\documentclass[journal=jctcce,manuscript=article,layout=preprint]{achemso}

\usepackage[T1]{fontenc} 
\usepackage{dcolumn}
\usepackage{bm}
\usepackage{color}
\usepackage{amsthm,amsmath,amssymb}
\usepackage{indentfirst}
\usepackage{caption}
\usepackage{hyperref}

\author{Seiji Ueno}
\email{se-ueno@hpc.co.jp}
\affiliation{HPC Systems Inc., Japan}
\affiliation{Department of Chemistry, Kyoto University, Kyoto, Japan}

\author{Yoshitaka Tanimura}
\email{tanimura@kuchem.kyoto-u.ac.jp}
\affiliation{Department of Chemistry, Kyoto University, Kyoto, Japan}

\title{Modeling and simulating the excited-state dynamics of a system with condensed phases: A machine learning approach}

\begin{document}

\begin{tocentry}
  \includegraphics[width=3.25in,height=1.75in]{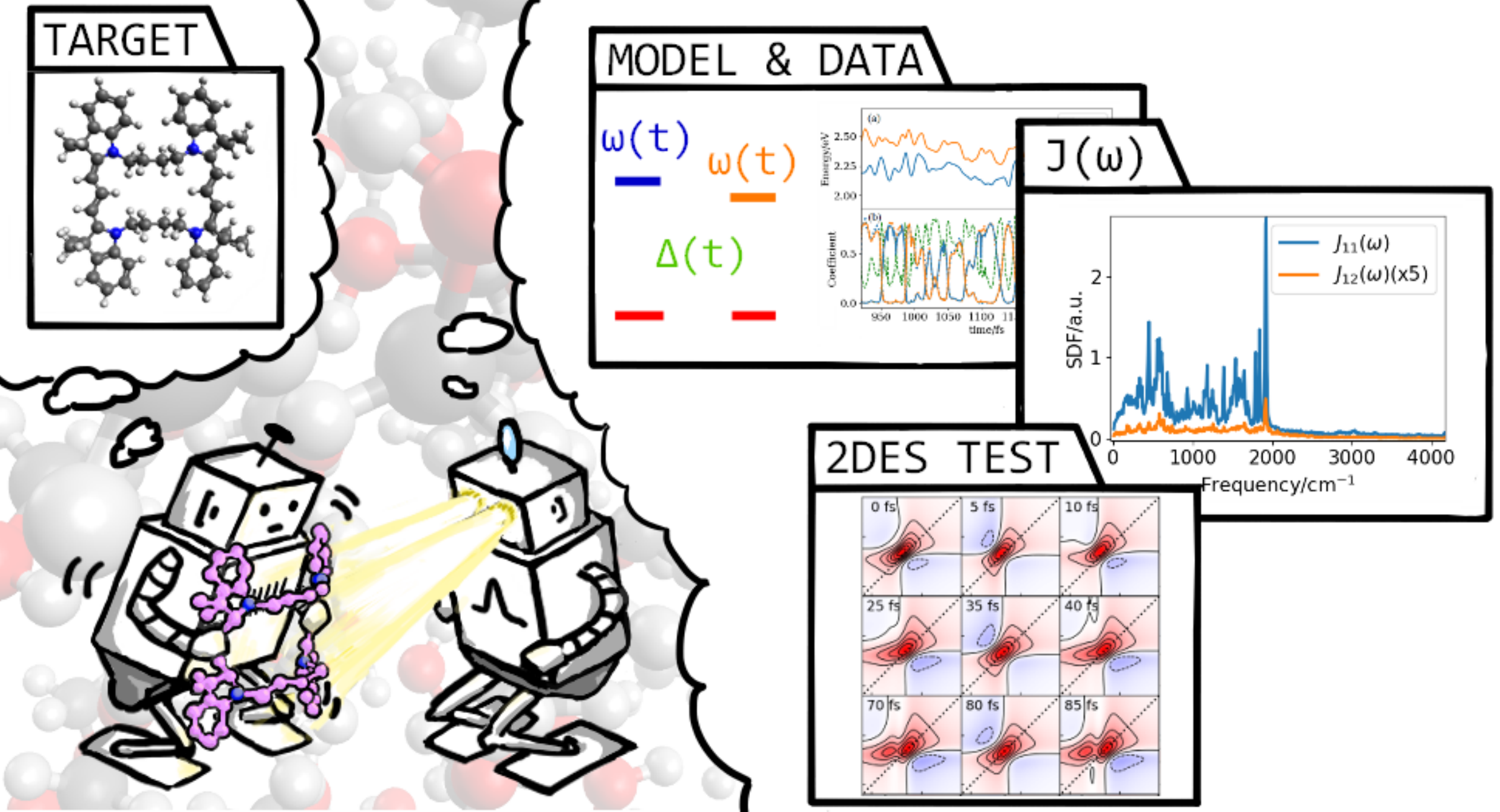}
\end{tocentry}

\begin{abstract}
Simulating the irreversible quantum dynamics of exciton and electron transfer problems poses a nontrivial challenge.
Because the irreversibility of the system dynamics is a result of quantum thermal activation and dissipation caused by the surrounding environment,
 it is necessary to include infinite environmental degrees of freedom in the simulation.
Because the capabilities of full quantum dynamics simulations that include the surrounding molecular degrees of freedom are limited, employing a system-bath model is a practical approach.
In such a model, the dynamics of excitons or electrons are described by a system Hamiltonian,
 while the other degrees of freedom that arise from the environmental molecules are described by a harmonic oscillator bath (HOB) and system-bath interaction parameters.
By extending on a previous study of molecular liquids [J. Chem. Theory Comput. 2020, 16, 2099],
 here we construct a system-bath model for exciton and electron transfer problems by means of a machine learning approach.
We determine both the system and system-bath interaction parameters, including the spectral distribution of the bath,
 using the electronic excitation energies obtained from a quantum mechanics/molecular mechanics (QM/MM) simulation that is conducted as a function of time.
Using the analytical expressions of optical response functions, we calculate linear and two-dimensional electronic (2DES) spectra for indocarbocyanine dimers in methanol.
From these results, we demonstrate the capability of our approach to elucidate the nonequilibrium exciton dynamics of a quantum system in a nonintuitive manner.
\end{abstract}


\section{INTRODUCTION}
Quantum dynamics play a significant role in many chemical physics and biochemical physics problems.
Frequently studied problems of this kind include exciton and electron transfer processes\cite{Khun95,YangTonuOliverRev2015} 
 that are involved in photosynthetic systems,\cite{KramerAspu14,LeeCoker2016,KramerFMO2DLorentz,SchultenJCP2009,SchultenJCP2011,SchultenJCP2012,SchultenFMO, Ishizaki2009,Renger2005,Renger2012,Renger2015, Renger2017,Valkunas2017, IshizakiJCP15, Nov2011, Nov2015, Mukamel2013, Renger06}
 electron transfer,\cite{Garg1985,Wolynes1987,Mukamel1988,TanakaJPSJ09,TanakaJCP10}
 DNA,\cite{SIM_MAKRI97,SIM2004,DijkstraNJP10DNA} and
photovoltaic systems.\cite{Gelinas2014, Thoss2015, Tamura2013, TamuraJPC2015, Prior2017,Mauro2020}
In these problems, the environments (baths), for example, proteins and solvents, play a central role;
 these baths are complex and strongly coupled to a molecular system of interest at finite temperatures.
Recent theoretical works have demonstrated that such systems and baths are quantum mechanically entangled (bathentanglement)
 and an understanding of these baths is essential to properly elucidate the quantum dynamics displayed by the system.\cite{TanimuraJPSJ06,YTpers}
For example, it has been shown that the optimal condition for excitation energy transfer in light-harvesting complexes is realized under non-Markovian system-bath interactions in a strong coupling regime,
 in which the noise correlation time of the bath is comparable to the time scale of the system dynamics.\cite{Ishizaki2009}
To conduct high-accuracy simulations with reduced computational costs, some approaches have utilized machine learning methods to develop models that reproduce open quantum dynamics,\cite{Pavlo2018ML,Hartmann2019NNQD,Flurin2020RNN,zheng_excitonic_2019}
 analyze two-dimensional spectroscopy images,\cite{rodriguez_machine_2019,namuduri_machine_2020}
 and estimate chemical properties for classical molecular dynamics.\cite{Smith2017ANI1,MLMD01, MLMD02, MLMD03, MLMD04}

Although an irreversibility of the system dynamics results from quantum thermal activation and dissipation caused by the surrounding environment,
 it is difficult to conduct a quantum molecular dynamics simulation that exhibits such a characteristic feature arising from macroscopic degrees of freedom.
Thus, we introduce a system-bath model in which the dynamics of excitons or electrons are described by a system Hamiltonian,
 while the other degrees of freedom that arise from environmental molecules are described by a harmonic-oscillator bath (HOB).
The HOB, whose distribution takes a Gaussian form, exhibits wide applicability in simulating bath effects, despite its simplicity;
 this is because the influence of the environment can, in many cases, be approximated by a Gaussian process due to the cumulative effect of the large number of environmental interactions.
In such a situation, the ordinary central limit theorem is applicable, and hence, the Gaussian distribution function is appropriate.\cite{TanimuraJPSJ06, Kampen81}
The distinctive features of the HOB model are determined by the spectral distribution function (SDF) of the coupling strength between the system and the bath oscillators for various frequency values.
By choosing the appropriate form of the SDF, the properties of the bath can be adjusted to represent various environments consisting of, for example, solid-state materials\cite{LipenHolns2015,ReichmanHolns2019}
 and protein molecules.\cite{KramerAspu14,LeeCoker2016,KramerFMO2DLorentz}
Because the SDF can be different for different forms of a system Hamiltonian and system-bath coupling,
 it is difficult to find an optimized Hamiltonian associated with an optimized SDF, in particular for a bath describing a fluctuation in site-site interaction energy.

In a previous study\cite{Ueno2020}, we employed a machine learning approach to construct a system-bath model for the intermolecular and intramolecular modes of molecular liquids
 using atomic trajectories obtained from molecular dynamics (MD) simulations.
In this study, we extend the previous approach to investigate an exciton or electron transfer problem that is characterized
 by electronic states embedded in the molecular environment using quantum mechanics/molecular mechanics (QM/MM) calculations to determine the atomic coordinates of molecules.
In particular, we focus on the exciton transfer process of the photosynthesis antenna system to investigate
 how natural systems can realize such highly efficient yields, presumably by manipulating quantum mechanical processes.
As a demonstration, we consider a molecular dimer made of two dipole-coupled dye monomers as a model system that is often studied experimentally and theoretically\cite{Dwayne2014,tempelaar_laser-limited_2016,duan_origin_2015}.
Then, we construct a model Hamiltonian of an indocarbocyanine dimer compound.\cite{Dwayne2014}
The accuracy of this model is examined by calculating linear and two-dimensional electronic spectra.

This paper is organized as follows. In Section 2, we introduce a model that can be used for either exciton or electron transfer and is coupled to a harmonic heat bath.
We then describe the machine learning approach that we use to determine the system parameters, the system-bath interactions, and the SDFs on the basis of QM/MM simulations.
In Section 3, we present results for an indocarbocyanine dimer model constructed from the analysis of QM/MM trajectories.
Linear absorption and two-dimensional spectra are calculated from analytical linear and nonlinear response functional expressions.
Section 6 is devoted to concluding remarks.

\section{THEORY}
\label{sec:Theory}
\subsection{Hamiltonian}
\label{subsec:Hamiltonian}

We consider the situations in which an exciton or electron transfer system interacts with molecular environments that give rise to dissipation and fluctuation in the system.
The Hamiltonian of the system is expressed as
\begin{equation}
\hat H_{S} = \sum _{j} \hbar \omega_j | j \rangle\langle j | + \sum _{j\ne k} \hbar \Delta_{jk}| j \rangle \langle k|,
\label{eq:system_hamiltonian}
\end{equation}
where the $j$th exciton or electron states with energies $\hbar \omega_j$ are represented by bra and ket vectors as  $| j \rangle$ and $\langle j |$.
The interaction energy between the $j$th and $k$th states is described by $\hbar \Delta_{jk}$.
In our model, each state is coupled to a different molecular environment (labeled as $a$) that is treated as $N_a$ harmonic oscillators.
The total Hamiltonian is then given by
\begin{align}
\label{eq:total_hamiltonian}
H_{tot} = & H_\mathrm{S} - \sum_{a} \sum_{l=1}^{N_{a}}\alpha_l^{a} \hat V^{a} \hat x_l^{a} \nonumber \\
 & +\sum_{a} \sum_{l=1}^{N_{a}} \left[ \frac{(\hat p_l^{a})^2 }{2m_l^{a} } + \frac{1}{2}m_l^{a} (\omega _l^{a})^2 (\hat x_l^{a})^2 \right],
\end{align}
where the momentum, position, mass, and frequency of the $l$th oscillator in the $a$th bath are given by $\hat{p}_{l}^{a}$, $\hat{x}_{l}^{a}$, $m_{l}^{a}$, and $\omega_{l}^{a}$, respectively.
The system part of the system-bath interaction is expressed as
\begin{equation}
\hat V^{a}= \sum _{j, k} V_{jk}^{a} | j \rangle\langle k |,
\label{eq:sys_interaction}
\end{equation}
where $V_{jk}^{a}$ is the coupling constant for the $a$th bath between the $j$ and $k$ states.
The $a$th heat bath can be characterized by the spectral distribution function (SDF), defined as
\begin{equation}
  J_a (\omega) \equiv \sum_{l=1}^{N_{a}}\frac{\hbar (\alpha_{l}^a)^2}{2m_{l}^a \omega_{l}^a } \delta(\omega-\omega_{l}^a),
  \label{eq:J_wgeneral}
\end{equation}
and the inverse temperature is $\beta \equiv 1/k_{\mathrm{B}}T$, where $k_\mathrm{B}$ is the Boltzmann constant.
Various environments, for example, those consisting of nanostructured materials, solvents, and protein molecules, can be modeled by adjusting the form of the SDF.
For the heat bath to act as an unlimited heat source possessing an infinite heat capacity,
 the number of heat-bath oscillators $N_a$ is effectively made infinitely large by replacing $J_a (\omega)$ with a continuous distribution.
The above model has been frequently used in the analysis of photosynthetic systems,\cite{SchultenJCP2009,SchultenJCP2011,SchultenJCP2012, Ishizaki2009,SchultenFMO,Renger2005,Renger2012,Renger2015, Renger2017,Valkunas2017, IshizakiJCP15, Nov2011, Nov2015, Mukamel2013, Renger06}
 electron transfer,\cite{Garg1985,Wolynes1987,Mukamel1988,TanakaJPSJ09,TanakaJCP10}
DNA,\cite{SIM_MAKRI97,SIM2004,DijkstraNJP10DNA} and solar battery systems.\cite{Gelinas2014, Thoss2015, Tamura2013, TamuraJPC2015, Prior2017,Mauro2020}

\subsection{Learning data: QM/MM simulations}
\label{subsec:ClassicalMD}
We next consider the pigments in a molecular system, whose electric excitation or exciton states are described by Eq.  \eqref{eq:system_hamiltonian}.
The electric states of the pigments depend on the configurations of the surrounding atoms at time $t$.
The time evolution of the excited states of the system and environmental molecules are described by QM/MM simulations.
Because our goal in constructing a system-bath model is to perform a full quantum simulation of the entire system,
 we should use quantum molecular dynamics (MD) simulations to provide data on the basis of all atomic coordinates.
In practice, however, it is impossible to consider large environmental degrees of freedom accurately from a quantum mechanical perspective.
Fortunately, we expect that we already have reasonable SDFs for quantum simulation, even though we evaluated them using the classical MD simulation.
Such evaluations were conducted utilizing an ensemble of molecular trajectories that exhibit a Gaussian distribution in which the difference between the quantum and classical trajectories is expected to be minor.
Further, the dynamics of harmonic oscillators are identical in both the classical and harmonic cases
 because both the classical and quantum Liouvillian for the $l$th oscillator in the $a$th bath are expressed as
 ${\hat L}_l^a =-({p_l^a}/{m_l^a})({\partial }/{\partial x_l^a})-m (\omega_l^a)^2 )({\partial }/{\partial p_l^a})$.
We thus use the classical MD simulation technique to acquire the atomic coordinates of the pigments and the molecular environment.
We then conduct quantum chemistry calculations to obtain the desired electronic states, typically the highest occupied molecular orbital (HOMO) and lowest unoccupied molecular orbital (LUMO) states of the pigments as a function of time.
The excited energy of the $j$th pigment is denoted by $\epsilon_{jj}(t)$, and the interaction energy between the $j$th and $k$th pigments that includes the bath-induced fluctuation is denoted by $\epsilon_{jk}(t)$;
 these values can be obtained using any kind of numerical program for quantum chemistry calculations.
If the main system is too large to enable evaluation of all electronic states, we evaluate the site energy $\epsilon_{jj}(t)$ and the interaction energy $\epsilon_{jk}(t)$ separately.
From the calculated $\epsilon_{jj}(t)$ and $\epsilon_{jk}(t)$,
 we evaluate the system-bath coupling strength in $\hat V_{jk}^{a}$ and its SDF, in addition to the excitation energy $\hbar \omega_j$ and the interaction energy $\Delta_{jk}$, based on the machine learning approach.

While the SDFs evaluated based on the MD simulations are temperature dependent, the SDFs for the HOB are temperature independent;
 we therefore eliminate the temperature dependence of optimized parameters, assuming that the sampled MD trajectories exhibit canonical ensembles at finite temperatures.

\subsection{Machine Learning}
\label{subsec:MachineLearing}
For $n$ exciton or electronic excitation sites, we express the simulated data in terms of $\epsilon_{jk}(t)$
 when describing the excited and site-site interaction energies of interest obtained from the QM/MM simulation.
The learning Hamiltonian is then expressed as
\begin{equation}
  \label{eq:hamiltonian_matrix}
  H(t) = \sum _{j, k=1}^n \epsilon_{jk}(t) | j \rangle \langle k|.
 \end{equation}
We then attempt to reproduce the trajectories of $\epsilon_{jk}(t)$
 for the total Hamiltonian, Eq. \eqref{eq:total_hamiltonian}, with Eqs. \eqref{eq:system_hamiltonian} and \eqref{eq:sys_interaction}.
Although the system-bath model considers an infinite number of degrees of freedom, here,
 we employ a finite number of bath oscillators to estimate the SDFs.
Then, the sampling used for machine leering training is considered the average of the classical bath oscillators
 for a certain selection of the system and system-bath parameters.
The site energy and interaction energy can be expressed as
\begin{equation}
  \epsilon_{jj}(t) =  \hbar\omega_j - \delta_{jj}(t)
\end{equation}
and
\begin{equation}
  \epsilon_{jk}(t) =  \hbar\Delta_{jk} - \delta_{jk}(t),
\end{equation}
respectively, where $\delta_{jk}(t)$ is expressed in terms of the linear function of the bath coordinates as
\begin{equation}
  \label{eq:ml_delta_define}
  \delta_{jk}(t) = \sum_a \alpha_{jk}^a x^a_{jk}(t).
\end{equation}
Here, the $a$th bath coordinate for the $jk$ site is described as a function of time as
\begin{equation}
  \label{eq:ml_xbath_define}
 x^a_{jk}(t) = A^a_{jk} \sin\left(\phi^a_{jk} + \omega^a_{jk} t \right),
\end{equation}
where $A^a_{jk}$ and $\phi^a_{jk}$ are the amplitude and phase of the $a$th bath oscillator for the $jk$ site, respectively.
The phase $\phi^a_{jk}$ is randomly chosen to avoid recursive oscillator motion.
Although we can consider such correlated modes separately by introducing additional baths,
 here, we assume that the influences of the individual bath modes are all independent and that the correlations between the fluctuations among different modes can be ignored.

From Eqs. \eqref{eq:ml_delta_define} and \eqref{eq:ml_xbath_define}, $\delta_{jk}(t)$ can be expressed as
\begin{equation}
  \delta_{jk}(t) = \sum_a c_{jk}^a 
\sin\left(\phi^a_{jk} + \omega^a_{jk} t \right),
\end{equation}
where 
\begin{equation}
  \label{eq:def_cjk}
c_{jk}^a = \alpha_{jk}^a A^a_{jk}
\end{equation}
and we treat the system-bath coupling parameters as the product of these two variables.
In the machine learning context, the bath parameters and the system-bath interactions are expressed as a set of latent variables, defined as
\begin{equation}
  \label{eq:ML_parameters}
  \theta = (\left\{\omega_{j}\right\},\left\{ \Delta_{jk}\right\},\left\{ c^a_{jk} \right\} ),
\end{equation}
where $\left\{... \right\}$ is the set of system and bath parameters.  
The trajectories of $\epsilon_{jj}(t)$ and $\epsilon_{jk}(t)$, obtained from the QM/MM calculations,
 are described as the vibrational motions of the pigment molecule and the surrounding molecules.
We then assume that the probability distribution of the pure state energy $\lambda_i$ is determined based on a Gaussian process and is described
 by a set of bath parameters $\alpha^a_{jk}$ and $\phi^a_{jk}$ by optimizing the probability distribution defined as
\begin{equation}
  P(\lambda_i\mid\theta) = \int \prod_{k,j,a} d\phi^a_{jk} P(\lambda_i\mid \theta; \phi^a_{jk})P(\phi^a_{jk}),
\end{equation}
which represents the marginalization of the phase of the oscillators $\phi^a_{jk}$ that is introduced to avoid trapping in a local minimal state due to the gradient method.
Here, $P(\phi^a_{jk})$ is the uniform distribution of $[0, 2\pi)$ and
\begin{equation}
 P(\lambda_i|\theta; \phi^a_{jk}) \propto \exp\left[-\sigma\left(\lambda_i - E_i \right)^2 \right],
\end{equation}
where $E_i \equiv E_i(\theta; \phi^a_{jk})$ is the predicted energy as a function of the parameter set $\theta$ and initial phase $\phi^a_{jk}$
 for the model Hamiltonian, Eq. \eqref{eq:hamiltonian_matrix}, and $\sigma$ is the error width.
Our goal in employing a machine learning method is to choose the optimal parameter set in $E_i(\theta; \phi^a_{jk})$ that maximizes the probability distribution for given data $\lambda_i$.
Among several optimization methods, we use the maximum likelihood method (MLE),
 where the loss function is expressed in terms of the negative log of the probability as
\begin{equation}
  L = \sum_i (\lambda_i - E_i)^2.
\end{equation}
To find the maximum value of $L$, we employ the Adam gradient method for optimization of the parameter set as
\begin{equation}
  \label{eq:Loss_gradient}
  \theta \leftarrow \theta + \gamma \frac{\partial L}{\partial \theta},
\end{equation}
where $\gamma$ is the learning rate.
In this way, we obtain the $J_{jk}$ element of the SDF for the $jk$ site.  
Because the energy distribution of each bath oscillator $E^a_{jk} = \frac{1}{2}m^a_{jk}\left(\omega^a_{jk}\right)^2$ is assumed to obey a canonical ensemble,
 the oscillator amplitude can be expressed as
\begin{equation}
  \left<A^a_{jk}\right> = \frac{1}{\sqrt{\pi\beta m^a_{jk}\left(\omega^a_{jk}\right)^2}}.
 \label{eq:avg_Ak}
\end{equation}
Integrating Eqs. \eqref{eq:def_cjk} and \eqref{eq:avg_Ak} into Eq. \eqref{eq:J_wgeneral}, we obtain
\begin{equation}
  \label{eq:J_fromML}
  J_{jk} (\omega) = \sum_{a=1}^{N_{a}}\frac{1}{2}\pi\beta\hbar\omega^a_{jk} (c_{jk}^a)^2 \delta(\omega-\omega_{jk}^a).
\end{equation}
Because $J_{jk}(\omega)$ rapidly changes over time in accordance with the structural changes in the pigment molecules and environments,
 we evaluate $J_{jk}(\omega)$ by averaging the different sample trajectories.
From a mathematical perspective, $c_{jk}$ is the frequency domain expression of the time domain data, and $J_{jk}(\omega)$ can be obtained
 by averaging the power spectra $c_{jk}^2$ using the Wiener-Khinchin theorem.

It should be noted that the absolute intensity of $J_{jk}(\omega)$ cannot be determined in the framework of the present study because,
 for simplicity, we do not evaluate the dipole moment of this complex material;
 we evaluate the intensity of $J_{jk}(\omega)$ from the width of the experimentally obtained linear absorption spectrum.

\section{Numerical demonstration}
\label{sec:numerical_demo}
\begin{figure}[htbp]
\includegraphics[width=0.4\columnwidth]{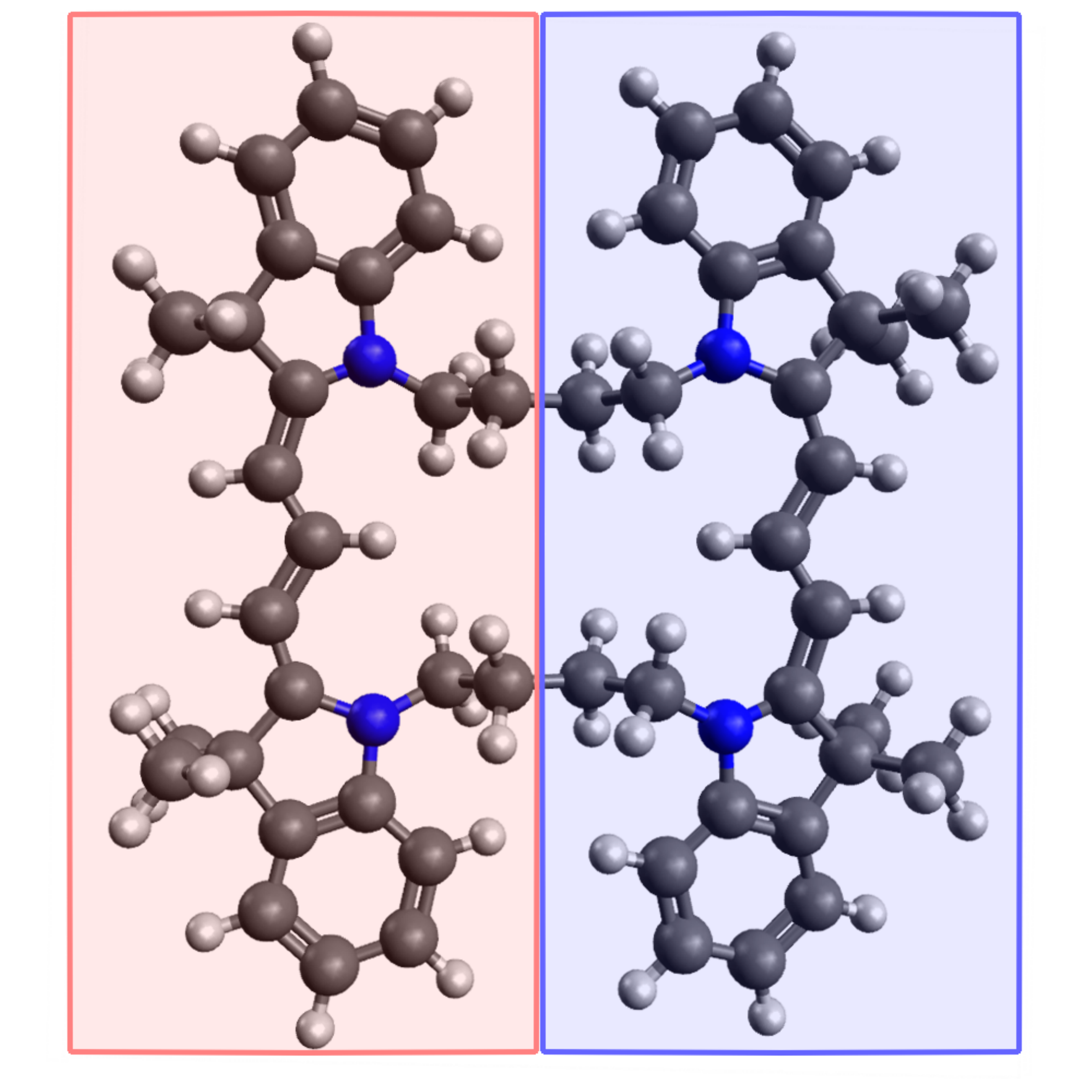}
\caption{ The molecular structure of the indocarbocyanine dimer.
Two pigments are connected by methylene chains.
The gray/blue/white atoms represent carbon/nitrogen/hydrogen, respectively.
The red square represents pigment 1, whereas the blue square represents pigment 2.
}
\label{fig:dimer_image}
\end{figure}

\subsection{Indocarbocyanine dimer}
We now demonstrate our numerical approach for a dimer of identical indocarbocyanine molecules.\cite{Dwayne2014}
Figure \ref{fig:dimer_image} displays the structure of the pigment molecule.
The ground and excited states of each pigment are expressed as $| 0\rangle_j$ and $| 1\rangle_j$ for $j = 1$ and 2, respectively.
The ground state energies are each set to zero.
The system Hamiltonian is then expressed as
\begin{align}
 \label{eq:Hamiltonian_dimer}
 \hat H = & \omega_{0}\left( | 1\rangle_1 {}_{1}\langle 1 | + | 1 \rangle_2 {}_{2}\langle 1 |\right) \nonumber \\
   & + \Delta\left(| 0 \rangle_1 {}_{2} \langle 1| + | 1 \rangle_1 {}_{2}\langle 0 |\right),
\end{align}
where $\omega_{0}$ is the excitation energy of a pigment and $\Delta$ is the interaction energy between the dimers.
By diagonalizing $H$, we obtained the eigenvalues $\omega_{k}$ for the $k=+$ and $-$ eigenstates of
 $| 1+ \rangle =(| 1\rangle_1| 0\rangle_2+| 0\rangle_1 | 1\rangle_2)/\sqrt{2}$ and  $| 1- \rangle =(| 1\rangle_1| 0\rangle_2-| 0\rangle_1 | 1\rangle_2)/\sqrt{2}$, respectively, as
\begin{equation}
  \omega_{\pm} = \omega_0 \pm \Delta.
\end{equation}
The excitation energy and interaction energy fluctuations as functions of time,
 arising from intramolecular motions of the pigment and intermolecular motions of surrounding molecules,
 are expressed as $\delta \omega_{\pm} (t)$ and $\delta\Delta(t)$, respectively.
These functions are evaluated based on the quantum chemistry calculations for given atomic trajectories of the entire molecular system determined by MD simulations.

In our model, because each exciton state is delocalized and the effects of the environmental modes are site specific,
 we employ an individual heat bath expressed as the sum of site-specific oscillators to describe the energy fluctuation at each exciton site. The distribution of the exciton-oscillator coupling strength is then evaluated based on the machine learning approach.
Although it is possible to introduce a global heat bath to induce low-frequency environmental modes that are coupled to multiple exciton states,
 we find that such effects are not significant in the present case.
Therefore, the excitation energy and interaction energy fluctuations are expressed as
\begin{align}
  \label{eq:delta_omega}
  \delta\omega_\pm(t) =  \sum_m^{1,2} w_{\pm,m}(t) \sum_a c^a_{\omega_0 m} \sin(\omega^a t + \phi^a_{\omega_0 m}),
\end{align}
and
\begin{align}
  \label{eq:delta_delta} 
  \delta\Delta(t) = w_{\Delta}(t)\sum_a c^a_{\Delta} \sin(\omega^a t + \phi^a_{\Delta}),
\end{align}
where $c^a_{b m}$ is the amplitude (scaled by $\alpha$, as described in Eq. \eqref{eq:def_cjk})
 and $\phi^a_{b m}$ is the initial phase of the $a$th oscillator for the state indices $b = 11$ (or $22$) and $12$.
We introduce the localization weight functions $w_{\pm ,m}(t)$ and $w_{\Delta}(t)$,
 as obtained from the diagonalization of the pigment-based Hamiltonian, expressed in Eq. \eqref{eq:Hamiltonian_dimer},
 to describe the pigment-specific environment effects in the delocalized exciton state representation.
These localization weight functions are evaluated based on the electronic states of the pigment $m = 1$ and $2$ established by the atomic orbitals (AO) obtained from quantum chemistry calculations.

Thus, the targeting eigenenergies to be described by the system-bath model, $\lambda_{\pm}(t; \theta)$, are expressed as
\begin{align}
  \lambda_{\pm}(t; \theta) = \omega_0 + \delta\omega_\pm(t)  \pm \left(\Delta + \delta\Delta(t)\right),
\end{align}
where $\theta$ is a set of parameters $\theta = \left(\omega_0, \left\{ c^a_{\pm ,m}\right\},\Delta,\left\{ c^a_{\Delta}\right\} \right)$.
As learning data, we compute the exciton energy $E_\pm(t)$, the molecular orbital (MO) coefficients for each exciton state,
 and wavefunctions (atomic orbital (AO) coefficients for each MO) from quantum chemistry calculations for the given atomic coordinates as a function of time.
Additionally, the movements of all atoms in the system are evaluated from the classical MD simulation.
Using these data, we optimize the set of parameters $\theta$.
To evaluate the weight function $w_{k, m}(t)$, we calculate the exciton and hole populations $p_{k,m}^{ex}(t)$ and $p_{k,m}^{h}(t)$ that are obtained as the summation of the absolute square of the AO coefficients,
 which are evaluated from the AO coefficients involved in the MO in pigment $m$ for excited state $k$.
The weight function is then evaluated as $w_{k,m}(t)=p_{k,m}^{ex}(t) p_{k,m}^{h}(t)$
 and $w_\Delta (t) = \sum_{k=\pm} \left( p_{k,1}^{h}(t) p_{k,2}^{ex}(t)+p_{k,2}^{h}(t) p_{k,1}^{ex}(t) \right)$.
As these definitions indicate, the exciton states are localized when $w_{\pm,m}$ is close to 1, whereas the exciton states are distributed among the pigments when $w_{\Delta}$ is close to 1.

To optimize the system and bath parameter set, we minimize the loss function
\begin{align}
  \label{eq:loss_function}
  L & = \sum_n\sum_t L^n(t) \nonumber \\
    & = \sum_n\sum_t \left[(\lambda_- (t; \theta^n) - E^n_{-}(t))^2 + (\lambda_+(t; \theta^n) - E^n_{+}(t))^2\right],
\end{align}
where $E^n_-(t)$ and $E^n_+(t)$ are the lowest ($\left|1-\right>$) and 2nd lowest ($\left|1+\right>$) excitation energies, and the index $n$ indicates the $n$th sample.
Using the MLE method, we optimize $c^a_{\omega_0 m}$ and $c^a_{\Delta}$  for each time series as a sample set.
To apply the machine learning algorithm, the time series of the tuple $(E^n_-(t), E^n_+(t), w^n_{k,m}(t))$ are regarded as the input feature variables.
In the indocarbocyanine case, the two pigments are symmetric, and the bath SDFs for each pigment are considered to be identical.
Therefore, we use the averaged value $c^a_{\omega_0} = (c^a_{\omega_0 1} + c^a_{\omega_0 2})/2$.
We then evaluate  $J_{jj}(\omega)  (j = 1, 2)$ and $J_{12}(\omega)$, namely, $J(\omega)$ for $\omega_0$ and $\Delta$, from $c^a_{\omega_0}$ and $c^a_\Delta$, respectively, using Eq. \eqref{eq:J_fromML}.

\subsection{Fourier-based approach versus machine learning approach}

A commonly used approach for evaluating the SDFs of $\epsilon_{ij}(t)$ utilizes the Fourier transformation of the autocorrelation function
 expressed as $\mathcal{F}\left[\left<\delta\epsilon_{ij}(0)\delta\epsilon_{ij}(t)\right>\right]$,
 where $\delta\epsilon_{ij}(t) \equiv \epsilon_{ij}(t) - \left<\epsilon_{ij}\right>$. 
In the actual calculation, the time series $\epsilon_{ij}^n(t)$, where $n$ is the sample index, is evaluated as the average of the autocorrelation function expressed as
\begin{equation}
C_{ij}(t) = \frac{1}{N}\sum_n\left<\delta\epsilon_{ij}^n(0)\delta\epsilon_{ij}^n(t)\right>,
\end{equation}
where $N$ is the total sample number.
We then obtain the SDF as
\begin{equation}
  \label{eq:gen_J_FT_C}
J_{ij}(\omega) = \mathcal{F}\left[C_{ij}(t)\right].
\end{equation}
Alternatively, using Wiener-Khinchin’s theorem for stationary random processes,
 we can obtain the SDF as an average of power spectrum $P^n_{ij}(\omega) = \left|\mathcal{F}\left[\epsilon_{ij}(t)\right]\right|^2$ as
\begin{equation}
  \label{eq:gen_J_avg_P}
J_{ij}(\omega) = \frac{1}{N}\sum_n P^n_{ij}(\omega).
\end{equation}
Although this Fourier-based approach is simple and straightforward,
 for the system-bath Hamiltonian, the obtained SDFs are not necessarily the optimal choice for describing the QM/MM data
 because the exciton and interaction energies are mutually dependent on each other; thus, $J_{ij}(\omega)$ and $J_{ik}(\omega)$ cannot be evaluated separately.
In the machine learning approach, however, it is possible to optimize not only $J_{ij}(\omega)$ and $J_{ik}(\omega)$ but also $\omega_0$ and $\Delta$
 without assuming explicit relationships between the SDFs and the system parameters.
Moreover, if necessary, we can introduce additional conditions for optimization of the SDFs and system parameters
 because we employed $w_{k,m}(t)$ to account for the effects of the indocarbocyanine dimer exciton localization.

\subsection{CALCULATION DETAILS}
\label{sec:Calculation}

\subsubsection*{Step 1: Classical MD}

We prepared a system consisting of an indocarbocyanine dimer molecule with 1024 methanol molecules as the solvent.
The classical MD simulations were carried out with the GROMACS software package. \cite{berendsen1995gromacs,abraham2015gromacs,bekker1993gromacs}
The conditions for preparation MD simulations were set as 1 atm and 300 K with an NPT ensemble.
The equilibrium MD run was carried out for 20 ps in an NVT ensemble followed by a sampling MD run for 5 ps in an NVE ensemble.
These equilibrium MD runs and sampling MD runs were repeated 100 times.
The entire MD simulation was performed with a time step of 0.1 fs.

\subsubsection*{Step 2: Data Preparation using Quantum Chemistry Calculations}

To obtain the sample trajectories of the excitation energies, we conducted ZINDO calculations\cite{ZINDO1973,ZINDO1991}
 and natural transition orbital analysis\cite{NTO2003} for a 1 fs period in one sample using the ORCA software package.\cite{neese2012orca}
We then obtained 100 $(E_-(t), E_+(t), w(t))$ samples that were 5 ps in length.

\subsubsection*{Step 3: Parameter Optimization for the Machine Learning Approach}

We arrange the data with 5 ps lengths obtained from step 2 according to the starting time in each of 175 steps.
We then extracted 604 trajectories containing 1000 data points in an interval of 4 fs.
These sampling data were used as the input feature values in the machine learning calculations.
To perform learning calculations, we developed Python codes using the TensorFlow library.\cite{tensorflow}
The training was performed with the learning rate $\alpha = 1\times 10^{-4}$ for the first 200 steps
 and then the rate was reduced to $\alpha = 1\times 10^{-5}$ for the next 200 steps.
The number of epochs was chosen to avoid the overfitting problem arising from the MLE that occurs with a gradient method. In the present case, this effect appears in the very law frequency region below 10 cm$^{-1}$ of $J(\omega)$ (see Appendix \ref{subsec:overfitting}). Because such slow dynamics of the environment are not important in the present exciton transfer problem, we avoided this effect by simply choosing a shorter epoch known as the early-stopping technique. To minimize the loss function, we employed the Adam algorithm.
The bath oscillator number $N$ is 600.
The frequency of the $a$th bath oscillator $\omega^a$ is  $a\Delta\omega$ for $a = 1, 2, \cdots, N$, where $\Delta\omega$ is approximately $8.3391 \mathrm{cm}^{-1}$.

The initial values of the target optimization variables for the SDF amplitudes were set as $c^a_{\omega_0 m}=1\times 10^{-5}$ and $c^a_{\Delta}=1\times 10^{-5}$,
 and the exciton and interaction energies were set as $\omega_0 = {(\left<E_+\right> +\left<E_-\right>)}/{2}$ and $\Delta = {(\left<E_+\right> - \left<E_-\right>)}/{2}$.
The initial phases $\phi^a_{b}$ were randomized 5 times for each series of samples.
The loss functions were averaged over each set of 64 samples as a minibatch, while the parameters were optimized for every minibatch.
For the 604 samples, each epoch contained 9 iterations.

\subsubsection*{Step 4: Calculations of Optical Spectra}

We assumed that the dipole operator for the indocarbocyanine dimer was given by
 $\hat \mu_1+\hat \mu_2 = \mu (| 0\rangle_1 {}_{1}\langle 1 |+| 1\rangle_1 {}_{1}\langle 0 | +| 0\rangle_2 {}_{2}\langle 1 |+| 1\rangle_2 {}_{2}\langle 0 |)$,
 which created a transition between the ground state  $|00 \rangle$ and the excitation states, $|1+ \rangle$, and $|11 \rangle$,
 while optical transitions from these states to the state $|1- \rangle$ were forbidden.
Thus, the optical transitions in the present system were modeled by a three-level system with eigenenergies of 0, $\Omega_+$, and $2\omega_0$.
This allowed us to apply analytical expressions of the linear and nonlinear response functions, as presented in appendix \ref{subsec:Spectroscopy}.
We then calculated the linear absorption and two-dimensional (2D) electronic spectroscopy signals using line-shape functions.

\subsection{RESULTS AND DISCUSSION}
\label{sec:Results}

\begin{figure}[htbp]
\includegraphics[width=0.6\columnwidth]{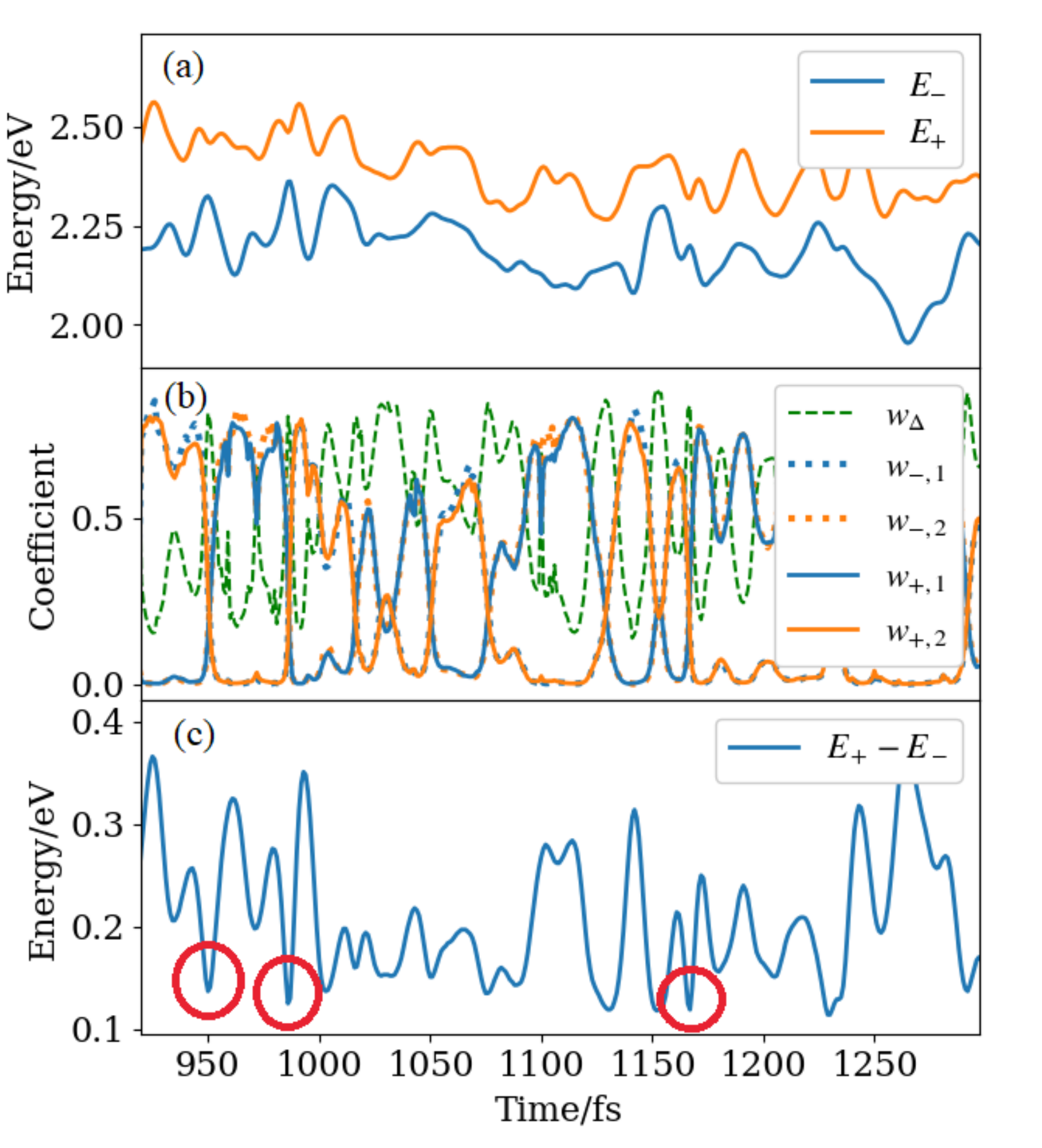}
\caption{
  Samples of the data used in learning calculations for (a) the excitation energy $E_{k}$ for $k=\pm$,
 (b) the weight functions $w_\Delta$ (green dashed curve), 
 $w_{k,m}(t)$ for pigments $m = 1$ (blue) and $m = 2$ (orange) for $k = +$ (solid line) and $k = -$ (dotted line), respectively. 
 The panel (c) is plotted the differences between the energy levels $E_\pm$ to illustrate the relationship between the energies and the weight functions.
}
\label{fig:data_sample}
\end{figure}

\begin{figure}[htbp]
\includegraphics[width=0.6\columnwidth]{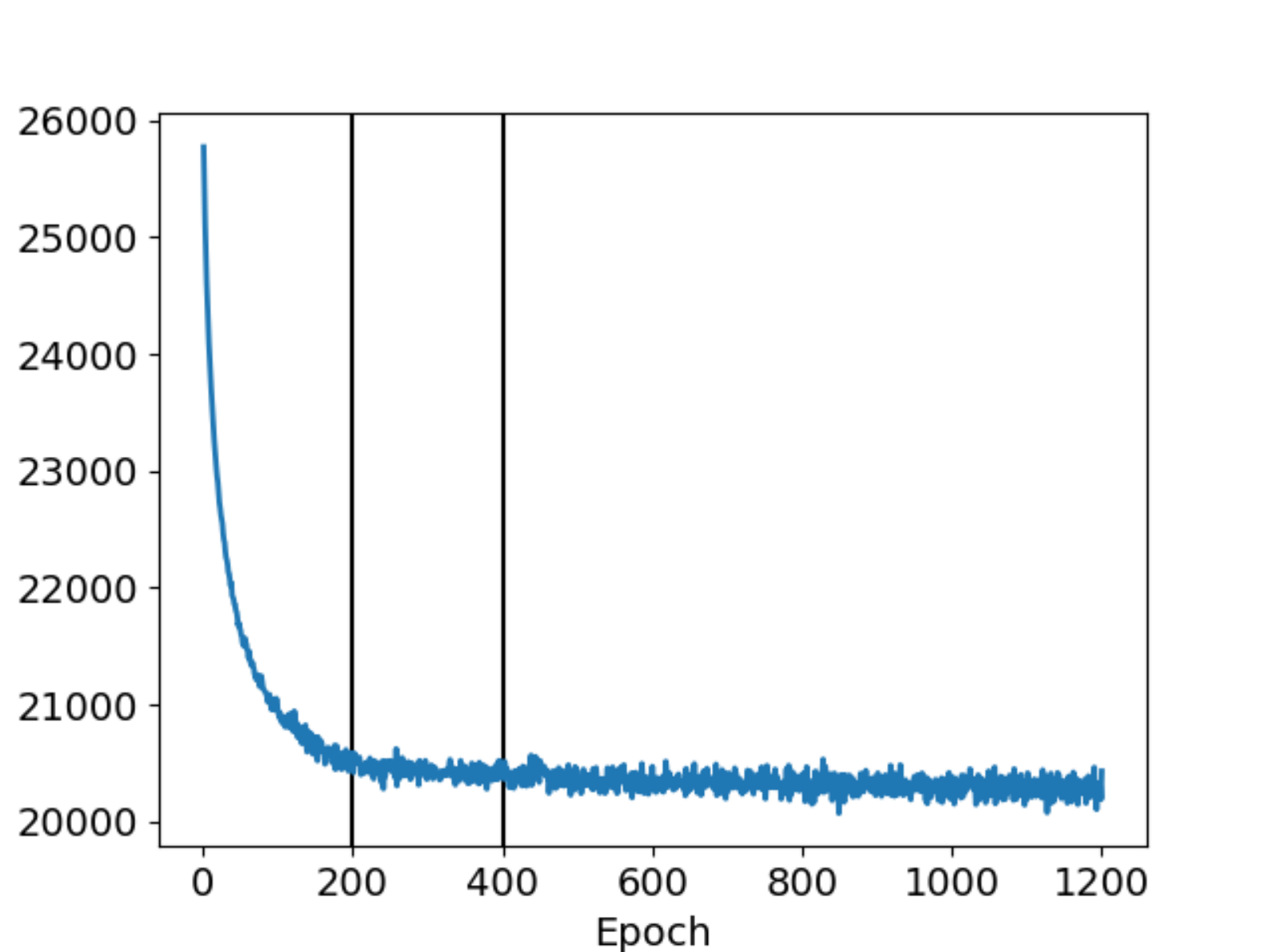}
\caption{The learning curve of the loss function for the indocarbocyanine dimer model.
The vertical line at epoch 200 indicates the epoch where the learning rate changed,
 and the vertical line at epoch 400 indicates early stopping.}
\label{fig:loss_function}
\end{figure}

\begin{figure}[htbp]
\includegraphics[width=0.8\columnwidth]{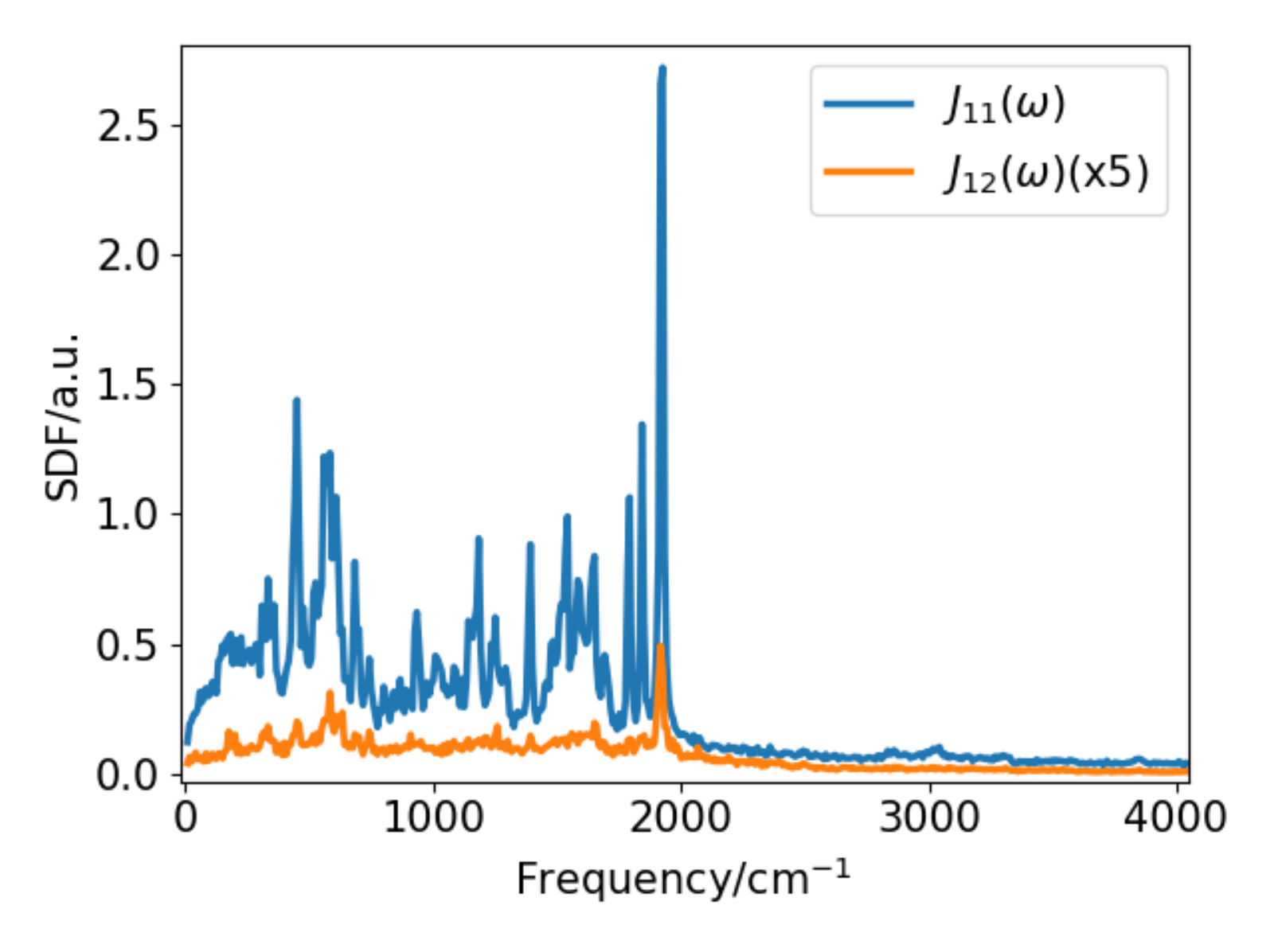}
\caption{ The SDFs of the indocarbocyanine dimer in the methanol environment for the exciton energy  $J_{11}(\omega)$ ($=J_{22}(\omega)$) (blue) and the interaction energy  $J_{12}(\omega)$ (orange)
 obtained with the machine learning approach.
}
\label{fig:J_omega}
\end{figure}

\begin{figure}[htbp]
\includegraphics[width=0.7\columnwidth]{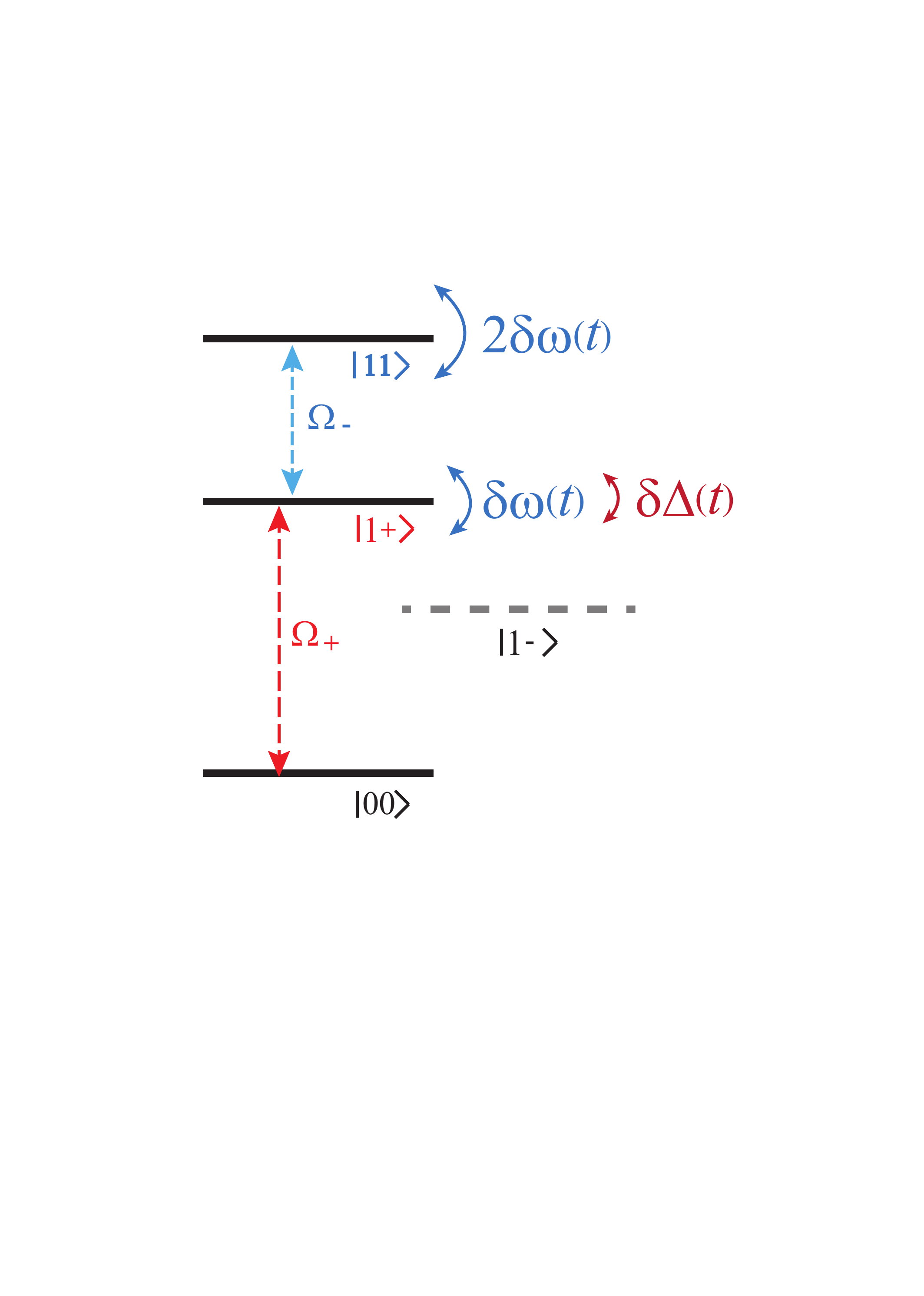}
\caption{Energy-level diagram for a dimer system that undergoes random fluctuations in the excited energy and coupling strength described by $\delta \omega(t)$ and $\delta \Delta(t)$, respectively.
For the description of pure dephasing, only the difference between the energies involved in the optical excitation is important:
 the frequency fluctuation between $|00 \rangle$ and $|1+ \rangle$ is given by $\delta \omega(t)+\delta \Delta(t)$,
 whereas that between $|1+ \rangle$  and $|11 \rangle$ is given by $\delta \omega(t)-\delta \Delta(t)$.
For perfectly uncorrelated fluctuations, we consider $\delta \omega(t)$ and $\delta \Delta(t)$ independently.
The dashed line represents the energy level of the forbidden state $|1 - \rangle$.
\label{triplet}
}
\end{figure}

Representative examples of the prepared dataset are plotted in Fig. \ref{fig:data_sample}.
The abrupt change in the exciton energies in Fig. \ref{fig:data_sample}(b) occurs due to the exciton transfer between pigments 1 and 2 that takes place in the time period of 10-100 fs.
As illustrated in Fig. \ref{fig:data_sample}(c), the difference in the exciton energies $E_+ - E_-$	exhibits minima in accordance with respect to the exciton transfer processes. 
As depicted by the red circles in Fig. \ref{fig:data_sample}(c), although such minimal points are significantly narrower and deeper than the minimal point caused by energetic fluctuation,
 it is difficult to separate the effects of exciton transfer from the energy fluctuation due to environmental motions.
By introducing the localization weight functions $w_{\pm,m}(t)$ and $w_{\Delta}(t)$ in Eqs. \eqref{eq:delta_omega} and \eqref{eq:delta_delta} to eliminate the effects of the nonenvironmental origin involved in the learning trajectories,
 we can stabilize and enhance the efficiency of the machine learning process.

In  Fig. \ref{fig:loss_function}, we depict the learning curve of the loss function, as defined in Eq. \eqref{eq:loss_function}.
Upon gathering random samplings of $\phi$, the loss function converged monotonically to a certain positive value, which demonstrated the efficiency of the present algorithm.
The initial parameter values of the excitation energy and the interaction energy were set as $\omega_0$ = 17736 cm$^{-1}$ and $\Delta$ = 1004 cm$^{-1}$,
 whereas the optimized values of the excitation energy and the interaction energy were given by $\omega_0$ = 17794 cm$^{-1}$ and $\Delta$ = 963 cm$^{-1}$,
 which are closer to the values that fit the experimentally obtained spectra.\cite{Dwayne2014} Here, to avoid overfitting problems,
we employed the early-stopping technique (see Appendix \ref{subsec:overfitting}).


In Fig. \ref{fig:J_omega}, we display the results of SDFs for the excitation energy $J_{11}(\omega)$ ($=J_{22}(\omega)$) and the interaction energy $J_{12}(\omega)$.
Various intermolecular modes below 2000 cm$^{-1}$ are observed as prominent sharp peaks near 450, 570, 1185, 1393, 1541, 1791, 1842, and 1923 cm$^{-1}$.
In the region above 2000 cm$^{-1}$, only two tiny peaks are observed at approximately 3000 cm$^{-1}$ and 3850 cm$^{-1}$.
The normal mode analysis (B3LYP/def-SV(P)) indicates that these peaks under 3300 cm$^{-1}$ arise from the intramolecular modes of the indocarbocyanine dimer,
 whereas the peak at 3850 cm$^{-1}$ arises from a molecular vibration of the solvent methanol molecules.
We found that each sharp peak can be fitted by the Brownian spectral distribution,\cite{TanakaJPSJ09,TanakaJCP10}
 whereas the broadened background peak in the range from 0 to 2000 cm$^{-1}$ corresponds to the intramolecular modes fitted by the Drude-Lorentz distribution.\cite{TanimruaJCP12}
The intensities of the peaks in $J_{12}(\omega)$ are considerably weaker than those in $J_{11}(\omega)$:
 only the peaks near 456, 562, 1840 and 1920 cm$^{-1}$ are identified.
As we expected, the intermolecular peak positions are governed by the classical MD simulation,
 whereas the heights of these peaks are predominately governed by the quantum chemistry calculation.

To verify the descriptions of the obtained SDFs and system parameters,
 we computed the linear absorption and two-dimensional electronic spectra (2DES),
 for the cases in which the experimentally obtained spectra were available.\cite{Dwayne2014}
In general, these spectra should be calculated in the framework of open quantum dynamics
 that considers the complex interactions between the exciton sites.
However, for demonstration purposes here, we employ the analytical expressions for response functions,
 ignoring the transitions to the state that are usually forbidden.
The details of these calculations are presented in Appendix A.

\begin{figure}[htbp]
\includegraphics[width=8cm]{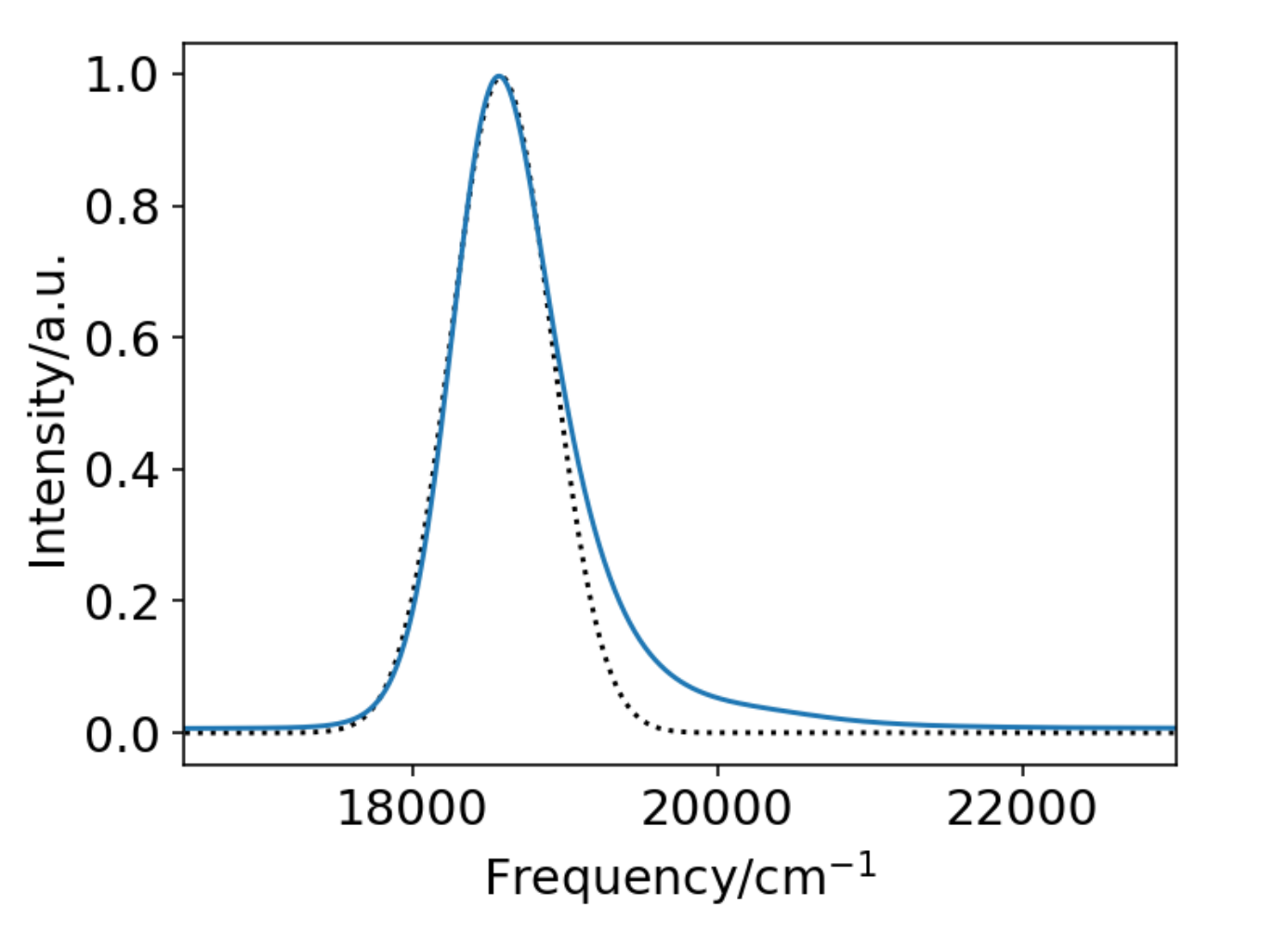}
\caption{
Linear absorption spectrum of an indocarbocyanine dimer, as calculated with Eqs. \eqref{eq:signal_1d}-\ref{eq:response_1d}
 and the line-shape function Eq. \eqref{LSF} for the system parameters and SDFs obtained with the machine learning approach.
The dotted line is the fitted Gaussian peak centered at $18583 \mathrm{cm}^{-1}$, indicating that the calculated peak is asymmetric due to the $0 - 1$ phonon transition near  $19500 \mathrm{cm}^{-1}$.
}
\label{fig:linear_absorb}
\end{figure}

The linear absorption spectrum calculated from Eqs. \eqref{eq:signal_1d} and \ref{eq:response_1d} is presented in Fig. \ref{fig:linear_absorb}.
Here, the calculated peak is fitted by the Gaussian function $\lambda\exp\left[-\left(({\omega - \omega_c})/{\gamma}\right)^2\right]$, where the amplitude, central frequency,
 and width are  $\lambda=351$, $\omega_c=18583 \mathrm{cm}^{-1}$, and $\gamma=464 \mathrm{cm}^{-1}$, respectively.
Note that we could not determine the absolute SDF intensities because, for simplicity, we did not calculate the amplitude of the dipole operator.
Here, we chose to use the intensity of $J_{11}(\omega)$ to fit the experimentally obtained signal.
As presented in Fig.  \ref{fig:linear_absorb}, we observe a single broadened absorption peak at $\omega_0+\Delta$
 corresponding to the transition between $|00 \rangle$ and $|1+ \rangle$, while the transition between $|00 \rangle$ and the state  $|1- \rangle$ is forbidden (see Fig.  \ref{triplet}).
Although the experimentally observed linear absorption spectrum exhibits a $0-1$ phonon sideband peak near $\omega=19500 \mathrm{cm}^{-1}$,
 here, we observe this phenomenon only as an asymmetry of the Gaussian peak in the high-frequency region.

\begin{figure}
\includegraphics[width=0.8\columnwidth]{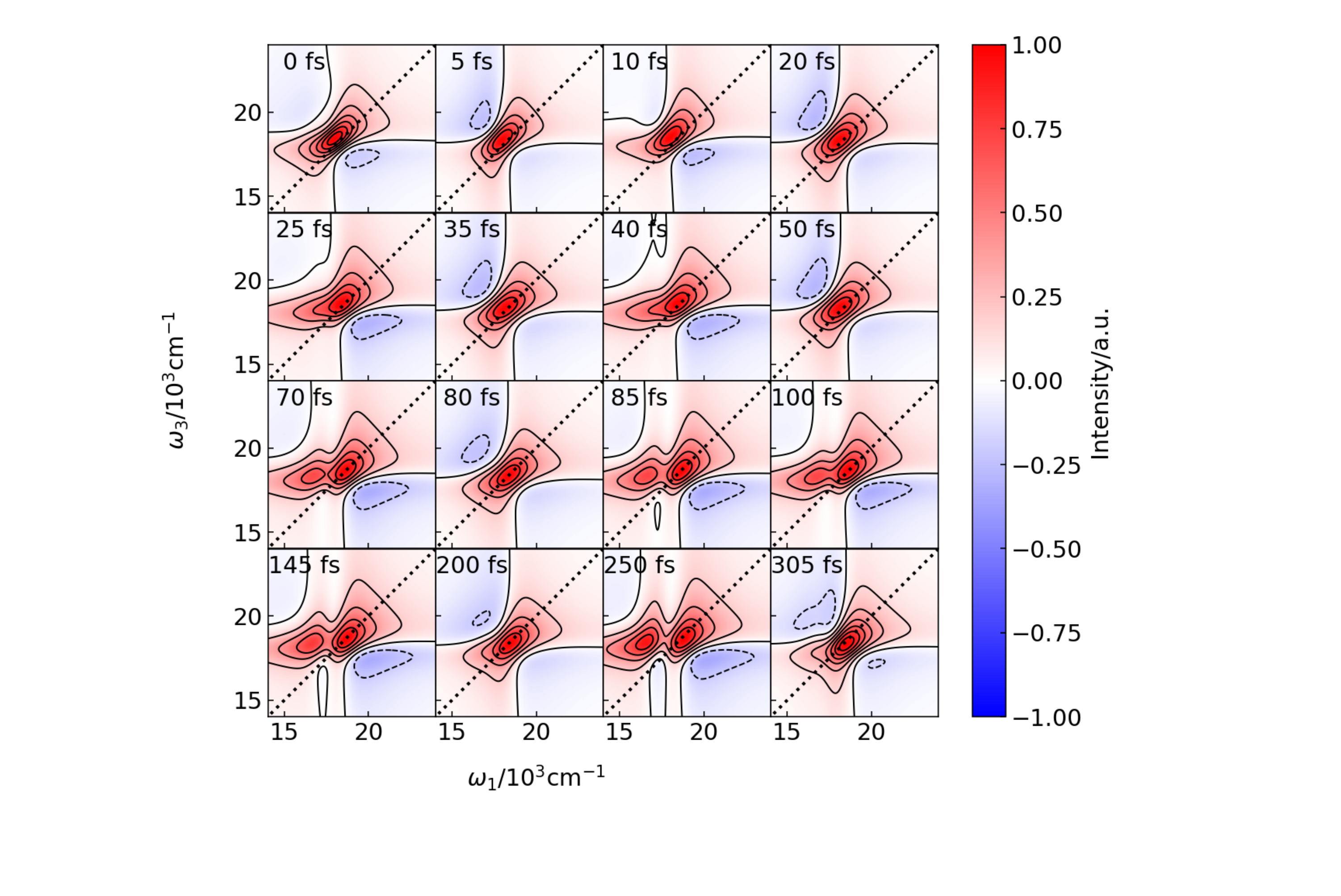}
\caption{
 2DES of an indocarbocyanine dimer, as calculated with Eqs. \eqref{resp1} -\ref{resp3} and the line-shape function Eq. \eqref{LSF}
  for the system parameters and SDFs obtained with the machine learning approach.
The waiting time $t_2$ for each signal is displayed at the top left of each panel.
The peak intensity of the signal was normalized for each $t_2$.
The waiting time $t_2$ was chosen to illustrate the maximal/minimal points of the oscillating feature of the peak elongation (see text).
}
\label{fig:2DES_sig}
\end{figure}

The 2D correlation electronic spectra calculated using the analytical expressions of the response function  (Eqs. \eqref{resp1} -\ref{resp3}) \cite{TanimuraMukamelJCP95}
are presented in Fig \ref{fig:2DES_sig}.
At $t_2 = 0$ fs, only one peak stretched near the  $\omega_1 = \omega_3$ line, arising from the  $|00\rangle \rightarrow |1+\rangle$ transition, is observed.
At $t_2 = 10$, $25$, and $40$ fs, the peak is elongated in the low-frequency $\omega_1$ direction due to a shift in the eigenenergy caused by the heat-bath-induced exciton-exciton interaction described by $J_{12}(\omega)$.
Because the system-bath interaction we considered here is non-Markovian and its effects appear only after a period longer than the inverse correlation time of noise, we do not observe such heat-bath effects for a small $t_2$.
Then, at approximately $t_2 = 70$ fs, the off-diagonal peak near  $(\omega_1, \omega_3) = (17800, 20000)$, in units of cm$^{-1}$,
 corresponding to the transition between  $|1+\rangle \rightarrow |11\rangle$ is observed,
 whereas the peak along the $\omega_1 = \omega_3$ line shifts to $(\omega_1, \omega_3) = (21000, 20000)$
 due to the transition between $|00\rangle \rightarrow |1+\rangle$ that arises from the exciton-exciton interaction described by $\Delta$ and $J_{12}(\omega)$.
As $t_2$ increases, the intensities of these two peaks oscillate as a result of the population transitions
 among $|10\rangle$, $|1+\rangle$, and $|11\rangle$ caused by $\Delta$ and $J_{12}(\omega)$.
This phenomenon was also observed experimentally.\cite{Dwayne2014}
The appearance of this oscillatory feature at a finite period in $t_2$ indicates the importance of the off-diagonal heat bath,
 whose modeling is not easy in the framework of the existing approach.

While the off-diagonal peak still exhibits oscillatory motion at $t_2 \ge 100$ fs, the peak profile gradually elongates in the $\omega_1 = \omega_3$ direction due to the inhomogeneous broadening that arises from the diagonal bath modulation described by $J_{11}(\omega)$ and $J_{22}(\omega)$.\cite{TanimuraJPSJ06}

\section{CONCLUSION}
\label{sec:conclusion}
We introduced a machine learning approach for constructing a model that can be used to analyze the dynamics of exciton or electron transfer processes in a complex environment
 on the basis of considering the energy eigenstates evaluated from QM/MM simulations as functions of time.
The key feature of the present study is the system-bath model,
 in which the primary exciton/electron dynamics are described by a system Hamiltonian expressed in terms of discretized energy states,
 while the other degrees of freedom are described by harmonic heat baths that are characterized by SDFs.
An optimized system-bath Hamiltonian obtained from the machine learning approach allows us
 to conduct time-irreversible quantum simulations that are not feasible with a full quantum MD simulation approach.

Here, we demonstrated the above features by calculating linear and nonlinear optical spectra for the indocarbocyanine dimer system in a methanol environment
 in which the quantum entanglement between the system and bath plays a central role.\cite{TanimuraJPSJ06,YTpers}
The calculated results can be used to explain the experimental results reasonably well;
 we found that the heat bath plays a key role in describing the exciton transfer process for the exciton-exciton interaction in this system.
Although here we ignore the transitions to the state that are usually forbidden due to an applicability of the analytical expression, if necessary,
 we can explicitly consider such transitions using the HEOM formalism.\cite{Mauro2020,TanimuraJPSJ06,YTpers,Y.Tanimura.JCP.2014,Y.Tanimura.JCP.2015}

Finally, we briefly discuss possible extensions of this study.
As shown in a previous paper,\cite{Ueno2020} the machine learning approach can be applied to a system described by reaction coordinates,
 which is useful for investigating chemical reaction processes characterized by potential energy surfaces.
By combining the previous and present approaches, we can further investigate systems described by not only electronic states but also molecular configuration space,
 for example, photoisomerization,\cite{T.Ikeda.JCP.2017} molecular motor,\cite{Ikeda2019} and nonadiabatic transition problems,\cite{Ikeda2018} with frameworks based on the system-bath model.
In this way, we may construct a system-bath model for entire photosynthesis reaction processes consisting of photoexcitation,\cite{Khun95,YangTonuOliverRev2015} 
 exciton transfer,\cite{KramerAspu14,LeeCoker2016,KramerFMO2DLorentz,SchultenJCP2009,SchultenJCP2011,SchultenJCP2012,SchultenFMO, Ishizaki2009,Renger2005,Renger2012,Renger2015, Renger2017,Valkunas2017, IshizakiJCP15, Nov2011, Nov2015, Mukamel2013, Renger06}
 electron transfer,\cite{Garg1985,Wolynes1987,Mukamel1988,TanakaJPSJ09,TanakaJCP10} and proton transfer processes,\cite{Shi2009PT,Shi2011PT,Jianji2020}
 including conversion processes, such as exciton-coupled electron transfer\cite{Sakamoto2017}
 and electron coupled proton transfer processes.\cite{Jianji2021}

Further theoretical and computational efforts must be put forth that include providing learning data based on accurate and large quantum simulations, improving learning algorithms,
 and developing an accurate and efficient open quantum dynamics theory to treat a complex system-bath model.
We leave such additional endeavors to future studies in accordance with recent progress in theoretical techniques.

\begin{acknowledgement}
The authors are thankful to Professor Yuki Kurashige for helpful discussions concerning the QM/MM simulations for providing an indocarbocyanine dimer system.
Financial support from HPC Systems Inc. is acknowledged.
\end{acknowledgement}

\appendix
\renewcommand{\theequation}{A.\arabic{equation} }
\renewcommand{\thefigure}{\Alph{section}.\arabic{table}}
\setcounter{equation}{0}
\section{One-dimensional and two-dimensional spectra}
\label{subsec:Spectroscopy}
Linear and nonlinear optical spectra can be expressed in the Fourier transformation of the response functions.\cite{mukamel1999book}
In the present dimer case, we can analytically express the response functions in terms of a line-shape function including the contribution from an exciton-exciton interaction.\cite{TanimuraMukamelJCP95}

The linear absorption spectrum is given by\cite{TanimuraMukamelPRE}
\begin{equation}
\label{eq:signal_1d}
S(\omega) = \int_0^\infty dt e^{i\omega t}R^{(1)}(t) -c.c.,
\end{equation} 
where $R^{(1)}(t) = \left<\left[\mu(t), \mu(0)\right]\right>$ is the one-dimensional (1D) response function expressed in terms of the transition dipole moment $\mu(t)$.
For a coupled dimer system, the analytical expression for the response function for Eq. \eqref{eq:signal_1d} is expressed as
\begin{align}
\label{eq:response_1d}
R^{(1)}(t_1) =& \frac{i\mu}{\hbar}\exp[i\Omega_+t_{1}-g_-^{11} (t_{1})-g_-^{12}(t_{1})] \nonumber \\
             & - c.c.
\end{align}
where the line-shape function, $\mathop g\nolimits_ \pm^a  (t)$, for the SDF, $J_a(\omega ) $, with $a=11$ and 12 is given by
\begin{align}
\mathop g\nolimits_ \pm^a  (t) &\equiv   \mathop \int \nolimits_0^t d't\mathop \int \nolimits_0^{t'} dt\int \frac{{d\omega }}{{2\pi }} \nonumber \\ 
    &\times J_a(\omega   ) \left[ {\coth\left( {\frac{{\beta \hbar \omega  }}{2}} \right) \cos(\omega   t) \pm i \sin(\omega   t)} \right] ,
\label{LSF}
\end{align}

\begin{figure}
\includegraphics[width=0.5\columnwidth]{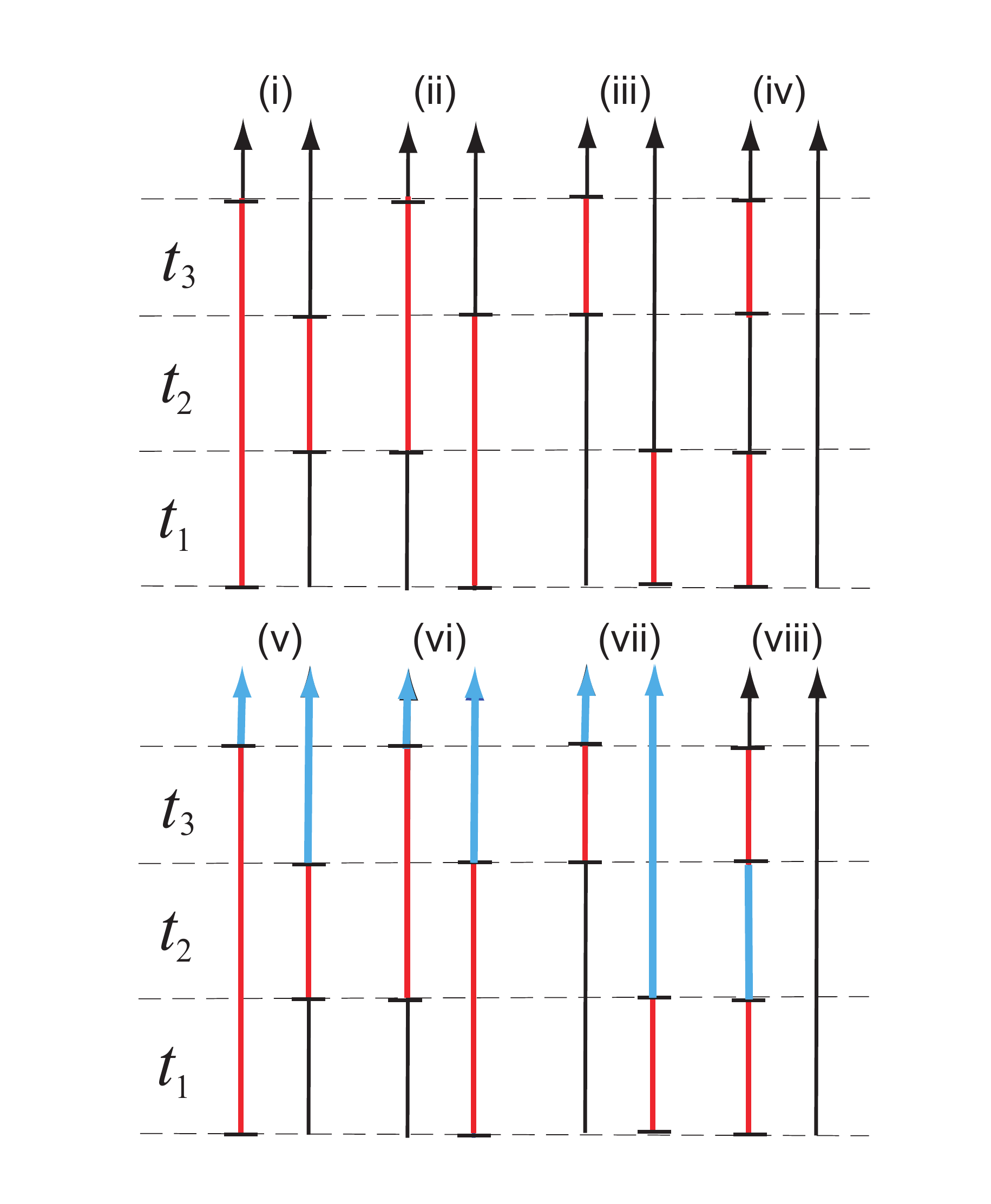}
\caption{Double-sided Feynman diagrams for the third-order response functions $R^{(3)} (t_3, t_2,t_1)$.
In each diagram, time increase from the bottom to the top, and $t_i$ represents the time intervals for the $i$th sequence between successive laser–system interactions.
The left line represents the time evolution of the ket, whereas the right line represents that of the bra.
The black, red and blue lines indicate that the system is in the  $|00 \rangle$, $|1+ \rangle$,  and $|11 \rangle$ states, respectively.
For these systems, the complex conjugate paths, which can be obtained by interchanging the ket and bra diagrams, are not shown here.
\label{Liouvillepath}
}
\end{figure}

For the coupled dimer system, the third-order response function is
\begin{align}
R^{(3)}(t_3, t_2,t_1) = \left<\left[\mu(t_3),\left[\mu(t_2)\left[\mu(t_1), \mu(0)\right]\right]\right]\right>
\end{align}
and can also be evaluated in the analytical form as\cite{TanimuraMukamelJCP95}
\begin{eqnarray}
\mathop R\nolimits^{(3)} (t_3, t_2,t_1) \equiv
 \frac{{i\mathop \mu \nolimits^4 }}{{\mathop \hbar \nolimits^3 }}\mathop \sum \limits_{\alpha  = 1}^8 \exp[\mathop Q\nolimits_\alpha  ({\rm{t)] }} -c.c.
\label{resp1}
\end{eqnarray}
where
\begin{equation}
\begin{split}
Q_1(t)=& -i\Omega_+t_{1}-i\Omega_+t_{3}                 \\& -f_1^1 (t_1,t_2,t_3) -f_1^2 (t_1,t_2,t_3), \\
Q_2(t)=&  i\Omega_+t_{1}-i\Omega_+t_{3}                 \\& -f_2^1 (t_1,t_2,t_3) -f_2^2 (t_1,t_2,t_3),\\
Q_3(t)=&  i\Omega_+t_{1}-i\Omega_+t_{3}                 \\& -f_3^1 (t_1,t_2,t_3) -f_3^2 (t_1,t_2,t_3),\\
Q_4(t)=& -i\Omega_+t_{1}-i\Omega_+t_{3}                 \\& -f_4^1 (t_1,t_2,t_3) -f_4^2 (t_1,t_2,t_3),\\
Q_5(t)=& -i\Omega_+t_{1}+i\Omega_-t_{3}                 \\& -f_5^1 (t_1,t_2,t_3) -f_5^2 (t_1,t_2,t_3),\\
Q_6(t)=&  i\Omega_+t_{1}+i\Omega_-t_{3}                 \\& -f_6^1 (t_1,t_2,t_3) -f_6^2 (t_1,t_2,t_3),\\
Q_7(t)=&  i\Omega_+t_{1}+2i\omega_0t_{2}+i\Omega_-t_{3} \\& -f_7^1 (t_1,t_2,t_3) -f_7^2 (t_1,t_2,t_3),\\ 
Q_8(t)=& -i\Omega_+t_{1}-2i\omega_0t_{2}-i\Omega_+t_{3} \\& -f_8^1 (t_1,t_2,t_3) -f_8^2 (t_1,t_2,t_3)
\label{resp2}
\end{split}
\end{equation}
with
\begin{equation}
\begin{split}
f_1^a &(t_1,t_2,t_3)= -g_-^a(t_{1})-g_+^a(t_{3}) \\ &-[g_+^a(t_{2})-g_+^a(t_{23})-g_-^a(t_{12})+g_-^a(t_{123})],\\
f_2^a &(t_1,t_2,t_3)= -g_+^a(t_{1})-g_+^a(t_{3}) \\ &+[g_-^a(t_{2})-g_-^a(t_{23})-g_+^a(t_{12})+g_+^a(t_{123})],\\
f_3^a &(t_1,t_2,t_3)= -g_+^a(t_{1})-g_-^a(t_{3}) \\ &+[g_+^a(t_{2})-g_+^a(t_{23})-g_+^a(t_{12})+g_+^a(t_{123})],\\
f_4^a &(t_1,t_2,t_3)= -g_-^a(t_{1})-g_-^a(t_{3}) \\ &-[g_-^a(t_{2})-g_-^a(t_{23})-g_-^a(t_{12})+g_-^a(t_{123})],\\ 
f_5^a &(t_1,t_2,t_3)= -g_-^a(t_{1})-g_+^a(t_{3}) \\ &-[g_+^a(t_{2})-g_+^a(t_{23})-g_-^a(t_{12})+g_-^a(t_{123})],\\
f_6^a &(t_1,t_2,t_3)= -g_+^a(t_{1})-g_+^a(t_{3}) \\ &+[g_-^a(t_{2})-g_-^a(t_{23})-g_+^a(t_{12})+g_+^a(t_{123})],\\
f_7^a &(t_1,t_2,t_3)= -g_+^a(t_{1})-g_-^a(t_{3}) \\ &+[g_+^a(t_{2})-g_+^a(t_{23})-g_+^a(t_{12})+g_+^a(t_{123})],\\
f_8^a &(t_1,t_2,t_3)= -g_-^a(t_{1})-g_-^a(t_{3}) \\ &-[g_-^a(t_{2})-g_-^a(t_{23})-g_-^a(t_{12})+g_-^a(t_{123})].
\label{resp3}
\end{split}
\end{equation}
Here, $t_{12}\equiv t_1+t_2$, $t_{23}\equiv t_2+t_3$, and $t_{123}\equiv t_1+t_2+t_3$.
Because the fluctuationis in the $|1+\rangle$ and  $|11\rangle$ states are described by $J_{11}(\omega) + J_{12}(\omega)$ and $J_{11}(\omega) + J_{22}(\omega) = 2J_{11}(\omega)$, respectively,
 the line-shape function $g_\pm^a(t)$ in Eq. \eqref{resp3} is now expressed as $g_\pm^1(t) = g_\pm^{11}(t) + g_\pm^{12}(t)$, and $g_\pm^2(t) = 2g_\pm^{11}(t)$.
By using third-order diagrams, the pump--probe spectrum and photon echo spectra are,
 for example, calculated from the $ Q_1(t)$, $ Q_4(t)$, and $ Q_5(t)$ elements,
 and the $ Q_2(t)$, $ Q_3(t)$, and $ Q_6(t)$ elements, respectively.

Although the change in the exciton population can be explored by pump-probe spectroscopy,
If we wish to investigate not only population dynamics but also system-bath coherence,
two-dimensional electronic correlation spectroscopy is a better choice.
This spectrum can be calculated from 
\begin{align} 
  I^{\mathrm{(corr)}}&(\omega _{3},t_{2},\omega _{1}) \nonumber \\
  &= I^{\mathrm{(NR)}}(\omega _{3},t_{2},\omega _{1}) + I^{\mathrm{(R)}}(\omega _{3},t_{2},\omega _{1}),
  \label{eq:I3}
\end{align}
where the non-rephasing and rephasing
parts of the signal are defined by
\begin{align} 
  I^{\mathrm{(NR)}}&(\omega _{3},t_{2},\omega _{1}) = \nonumber \\
  &\mathrm{Im}\int _{0}^{\infty }\mathrm{d}t_{3}\int _{0}^{\infty }\mathrm{d}t_{1}e^{i\omega _{3}t_{3}}e^{i\omega _{1}t_{1}}R^{(3)}(t_{3},t_{2},t_{1}),
  \label{eq:nonreph}
\intertext{and}
  I^{\mathrm{(R)}}&(\omega _{3},t_{2},\omega _{1}) = \nonumber\\
  & \mathrm{Im}\int _{0}^{\infty }\mathrm{d}t_{3}\int _{0}^{\infty }\mathrm{d}t_{1}e^{i\omega_{3}t_{3}}e^{-i\omega _{1}t_{1}}R^{(3)}(t_{3},t_{2},t_{1}).
  \label{eq:reph}
\end{align}

\section{Overfitting problem of MLE}
\label{subsec:overfitting}
\begin{figure}[htbp]
\includegraphics[width=0.8\columnwidth]{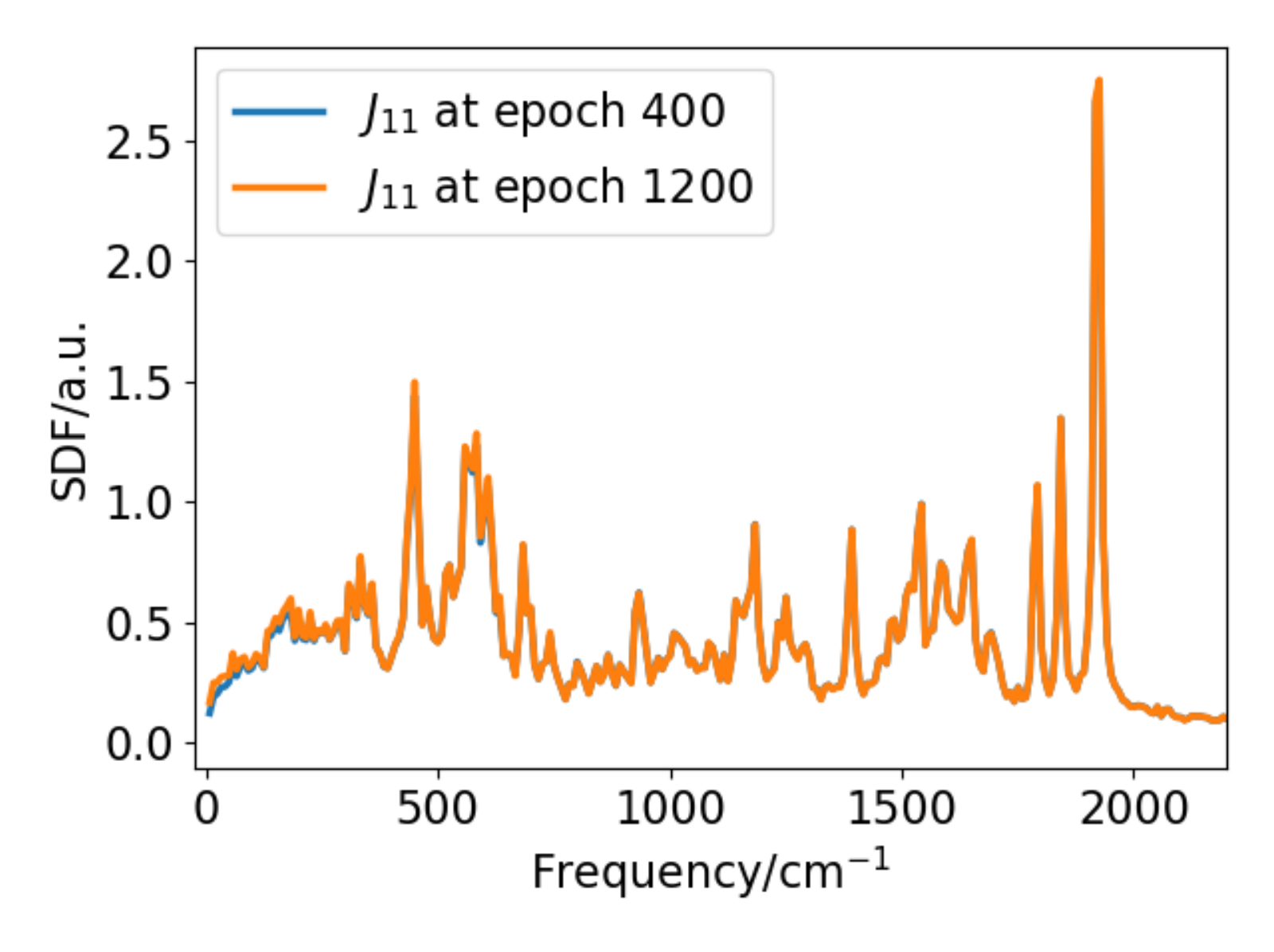}
\caption{
SDF $J_{11}(\omega)$ ($=J_{22}(\omega)$) obtained in the early-stopping case (blue) and overfitting case (orange).
}
\label{fig:J_omega_ratio}
\end{figure}
To illustrate the overfitting problem effect,
 we present the optimized results evaluated in the early-stopping case (epoch 400) and overfitting case (epoch 1200).
The learning curve of the loss function for these two cases is presented in Fig. \ref{fig:loss_function}.
As illustrated in Fig. \ref{fig:J_omega_ratio}, the results are similar and overlap everywhere except the low frequency region, $0 \sim 400 \mathrm{cm}^{-1}$.
The optimized system parameters in the early-stopping case are $\omega_0$ = 17794 cm$^{-1}$ and $\Delta$ = 963 cm$^{-1}$,
 whereas those in the overfitting case are $\omega_0$ = 17795 cm$^{-1}$ and $\Delta$ = 960 cm$^{-1}$.
These results indicate that the learning process works very well even in epoch 400.
We then found that the accuracy of the obtained SDF, particularly in the region below 400 $\mathrm{cm}^{-1}$, decreases for the case of epoch 1200,
 because the fluctuation of the loss function in a larger epoch period suppresses the convergence of the SDF in the low frequency region, as illustrated in Fig. \ref{fig:loss_function}.
This phenomenon is known as the overfitting (or overtraining) problem of the MLE for the gradient method.
We may avoid this problem by regularizing the model, for example, by adopting an L2 regularization, or by using the Bayesian inference method to account for the physical knowledge as a prior probability.\cite{bishop_pattern_2006}
Nevertheless, here we use an early-stopping method to simply reduce the numerical cost,
 because the low-frequency region $100 \le \omega \le 400 \mathrm{cm}^{-1}$ is no longer significant in the ultrafast dynamics of the exciton transfer problem,
 whereas the region below 10 $\mathrm{cm}^{-1}$ may alter the signal profile significantly due to the quantum thermal factor, $\coth\left(\beta\hbar\omega/2\right)$,
 in the line-shape function defined as Eq. \eqref{LSF}.

\clearpage

\bibliography{refs}

\providecommand{\latin}[1]{#1}
\makeatletter
\providecommand{\doi}
  {\begingroup\let\do\@makeother\dospecials
  \catcode`\{=1 \catcode`\}=2 \doi@aux}
\providecommand{\doi@aux}[1]{\endgroup\texttt{#1}}
\makeatother
\providecommand*\mcitethebibliography{\thebibliography}
\csname @ifundefined\endcsname{endmcitethebibliography}
  {\let\endmcitethebibliography\endthebibliography}{}
\begin{mcitethebibliography}{78}
\providecommand*\natexlab[1]{#1}
\providecommand*\mciteSetBstSublistMode[1]{}
\providecommand*\mciteSetBstMaxWidthForm[2]{}
\providecommand*\mciteBstWouldAddEndPuncttrue
  {\def\EndOfBibitem{\unskip.}}
\providecommand*\mciteBstWouldAddEndPunctfalse
  {\let\EndOfBibitem\relax}
\providecommand*\mciteSetBstMidEndSepPunct[3]{}
\providecommand*\mciteSetBstSublistLabelBeginEnd[3]{}
\providecommand*\EndOfBibitem{}
\mciteSetBstSublistMode{f}
\mciteSetBstMaxWidthForm{subitem}{(\alph{mcitesubitemcount})}
\mciteSetBstSublistLabelBeginEnd
  {\mcitemaxwidthsubitemform\space}
  {\relax}
  {\relax}

\bibitem[May and Kühn(2003)May, and Kühn]{Khun95}
May,~V.; Kühn,~O. \emph{Charge and Energy Transfer Dynamics in Molecular
  Systems}; Wiley-{VCH}, 2003\relax
\mciteBstWouldAddEndPuncttrue
\mciteSetBstMidEndSepPunct{\mcitedefaultmidpunct}
{\mcitedefaultendpunct}{\mcitedefaultseppunct}\relax
\EndOfBibitem
\bibitem[Schröter \latin{et~al.}(2015)Schröter, Ivanov, Schulze, Polyutov,
  Yan, Pullerits, and Kühn]{YangTonuOliverRev2015}
Schröter,~M.; Ivanov,~S.; Schulze,~J.; Polyutov,~S.; Yan,~Y.; Pullerits,~T.;
  Kühn,~O. Exciton–vibrational coupling in the dynamics and spectroscopy of
  Frenkel excitons in molecular aggregates. \emph{Physics Reports}
  \textbf{2015}, \emph{567}, 1--78\relax
\mciteBstWouldAddEndPuncttrue
\mciteSetBstMidEndSepPunct{\mcitedefaultmidpunct}
{\mcitedefaultendpunct}{\mcitedefaultseppunct}\relax
\EndOfBibitem
\bibitem[Kreisbeck \latin{et~al.}(2014)Kreisbeck, Kramer, and
  Aspuru-Guzik]{KramerAspu14}
Kreisbeck,~C.; Kramer,~T.; Aspuru-Guzik,~A. Scalable High-Performance Algorithm
  for the Simulation of Exciton Dynamics. Application to the Light-Harvesting
  Complex {II} in the Presence of Resonant Vibrational Modes. \emph{J. Chem.
  Theory Comput.} \textbf{2014}, \emph{10}, 4045--4054\relax
\mciteBstWouldAddEndPuncttrue
\mciteSetBstMidEndSepPunct{\mcitedefaultmidpunct}
{\mcitedefaultendpunct}{\mcitedefaultseppunct}\relax
\EndOfBibitem
\bibitem[Lee and Coker(2016)Lee, and Coker]{LeeCoker2016}
Lee,~M.~K.; Coker,~D.~F. Modeling Electronic-Nuclear Interactions for
  Excitation Energy Transfer Processes in Light-Harvesting Complexes. \emph{J.
  Phys. Chem. Lett.} \textbf{2016}, \emph{7}, 3171--3178\relax
\mciteBstWouldAddEndPuncttrue
\mciteSetBstMidEndSepPunct{\mcitedefaultmidpunct}
{\mcitedefaultendpunct}{\mcitedefaultseppunct}\relax
\EndOfBibitem
\bibitem[Kreisbeck and Kramer(2012)Kreisbeck, and Kramer]{KramerFMO2DLorentz}
Kreisbeck,~C.; Kramer,~T. Long-Lived Electronic Coherence in Dissipative
  Exciton Dynamics of Light-Harvesting Complexes. \emph{J. Phys. Chem. Lett.}
  \textbf{2012}, \emph{3}, 2828--2833\relax
\mciteBstWouldAddEndPuncttrue
\mciteSetBstMidEndSepPunct{\mcitedefaultmidpunct}
{\mcitedefaultendpunct}{\mcitedefaultseppunct}\relax
\EndOfBibitem
\bibitem[Strümpfer and Schulten(2009)Strümpfer, and
  Schulten]{SchultenJCP2009}
Strümpfer,~J.; Schulten,~K. Light harvesting complex {II} B850 excitation
  dynamics. \emph{J. Chem. Phys.} \textbf{2009}, \emph{131}, 225101\relax
\mciteBstWouldAddEndPuncttrue
\mciteSetBstMidEndSepPunct{\mcitedefaultmidpunct}
{\mcitedefaultendpunct}{\mcitedefaultseppunct}\relax
\EndOfBibitem
\bibitem[Strümpfer and Schulten(2011)Strümpfer, and
  Schulten]{SchultenJCP2011}
Strümpfer,~J.; Schulten,~K. The effect of correlated bath fluctuations on
  exciton transfer. \emph{J. Chem. Phys.} \textbf{2011}, \emph{134},
  095102\relax
\mciteBstWouldAddEndPuncttrue
\mciteSetBstMidEndSepPunct{\mcitedefaultmidpunct}
{\mcitedefaultendpunct}{\mcitedefaultseppunct}\relax
\EndOfBibitem
\bibitem[Strümpfer and Schulten(2012)Strümpfer, and
  Schulten]{SchultenJCP2012}
Strümpfer,~J.; Schulten,~K. Excited state dynamics in photosynthetic reaction
  center and light harvesting complex 1. \emph{J. Chem. Phys.} \textbf{2012},
  \emph{137}, 065101\relax
\mciteBstWouldAddEndPuncttrue
\mciteSetBstMidEndSepPunct{\mcitedefaultmidpunct}
{\mcitedefaultendpunct}{\mcitedefaultseppunct}\relax
\EndOfBibitem
\bibitem[Olbrich \latin{et~al.}(2011)Olbrich, Strümpfer, Schulten, and
  Kleinekathöfer]{SchultenFMO}
Olbrich,~C.; Strümpfer,~J.; Schulten,~K.; Kleinekathöfer,~U. Quest for
  Spatially Correlated Fluctuations in the {FMO} Light-Harvesting Complex.
  \emph{J. Phys. Chem. B} \textbf{2011}, \emph{115}, 758--764\relax
\mciteBstWouldAddEndPuncttrue
\mciteSetBstMidEndSepPunct{\mcitedefaultmidpunct}
{\mcitedefaultendpunct}{\mcitedefaultseppunct}\relax
\EndOfBibitem
\bibitem[Ishizaki and Fleming(2009)Ishizaki, and Fleming]{Ishizaki2009}
Ishizaki,~A.; Fleming,~G.~R. On the adequacy of the Redfield equation and
  related approaches to the study of quantum dynamics in electronic energy
  transfer. \emph{J. Chem. Phys.} \textbf{2009}, \emph{130}, 234110\relax
\mciteBstWouldAddEndPuncttrue
\mciteSetBstMidEndSepPunct{\mcitedefaultmidpunct}
{\mcitedefaultendpunct}{\mcitedefaultseppunct}\relax
\EndOfBibitem
\bibitem[Raszewski \latin{et~al.}(2005)Raszewski, Saenger, and
  Renger]{Renger2005}
Raszewski,~G.; Saenger,~W.; Renger,~T. Theory of Optical Spectra of Photosystem
  {II} Reaction Centers: Location of the Triplet State and the Identity of the
  Primary Electron Donor. \emph{Biophys. J.} \textbf{2005}, \emph{88},
  986--998\relax
\mciteBstWouldAddEndPuncttrue
\mciteSetBstMidEndSepPunct{\mcitedefaultmidpunct}
{\mcitedefaultendpunct}{\mcitedefaultseppunct}\relax
\EndOfBibitem
\bibitem[Renger \latin{et~al.}(2012)Renger, Klinger, Steinecker, Schmidt~am
  Busch, Numata, and Müh]{Renger2012}
Renger,~T.; Klinger,~A.; Steinecker,~F.; Schmidt~am Busch,~M.; Numata,~J.;
  Müh,~F. Normal Mode Analysis of the Spectral Density of the
  Fenna–Matthews–Olson Light-Harvesting Protein: How the Protein Dissipates
  the Excess Energy of Excitons. \emph{J. Phys. Chem. B} \textbf{2012},
  \emph{116}, 14565--14580\relax
\mciteBstWouldAddEndPuncttrue
\mciteSetBstMidEndSepPunct{\mcitedefaultmidpunct}
{\mcitedefaultendpunct}{\mcitedefaultseppunct}\relax
\EndOfBibitem
\bibitem[Müh \latin{et~al.}(2015)Müh, Plöckinger, Ortmayer, Schmidt~am
  Busch, Lindorfer, Adolphs, and Renger]{Renger2015}
Müh,~F.; Plöckinger,~M.; Ortmayer,~H.; Schmidt~am Busch,~M.; Lindorfer,~D.;
  Adolphs,~J.; Renger,~T. The quest for energy traps in the {CP}43 antenna of
  photosystem {II}. \emph{J. Photochem. Photobiol., B} \textbf{2015},
  \emph{152}, 286--300\relax
\mciteBstWouldAddEndPuncttrue
\mciteSetBstMidEndSepPunct{\mcitedefaultmidpunct}
{\mcitedefaultendpunct}{\mcitedefaultseppunct}\relax
\EndOfBibitem
\bibitem[Müh \latin{et~al.}(2017)Müh, Plöckinger, and Renger]{Renger2017}
Müh,~F.; Plöckinger,~M.; Renger,~T. Electrostatic Asymmetry in the Reaction
  Center of Photosystem {II}. \emph{J. Phys. Chem. Lett.} \textbf{2017},
  \emph{8}, 850--858\relax
\mciteBstWouldAddEndPuncttrue
\mciteSetBstMidEndSepPunct{\mcitedefaultmidpunct}
{\mcitedefaultendpunct}{\mcitedefaultseppunct}\relax
\EndOfBibitem
\bibitem[Gelzinis \latin{et~al.}(2017)Gelzinis, Abramavicius, Ogilvie, and
  Valkunas]{Valkunas2017}
Gelzinis,~A.; Abramavicius,~D.; Ogilvie,~J.~P.; Valkunas,~L. Spectroscopic
  properties of photosystem {II} reaction center revisited. \emph{J. Chem.
  Phys.} \textbf{2017}, \emph{147}, 115102\relax
\mciteBstWouldAddEndPuncttrue
\mciteSetBstMidEndSepPunct{\mcitedefaultmidpunct}
{\mcitedefaultendpunct}{\mcitedefaultseppunct}\relax
\EndOfBibitem
\bibitem[Fujihashi \latin{et~al.}(2015)Fujihashi, Fleming, and
  Ishizaki]{IshizakiJCP15}
Fujihashi,~Y.; Fleming,~G.~R.; Ishizaki,~A. Impact of environmentally induced
  fluctuations on quantum mechanically mixed electronic and vibrational pigment
  states in photosynthetic energy transfer and 2D electronic spectra. \emph{J.
  Chem. Phys.} \textbf{2015}, \emph{142}, 212403\relax
\mciteBstWouldAddEndPuncttrue
\mciteSetBstMidEndSepPunct{\mcitedefaultmidpunct}
{\mcitedefaultendpunct}{\mcitedefaultseppunct}\relax
\EndOfBibitem
\bibitem[Novoderezhkin \latin{et~al.}(2011)Novoderezhkin, Romero, Dekker, and
  van Grondelle]{Nov2011}
Novoderezhkin,~V.~I.; Romero,~E.; Dekker,~J.~P.; van Grondelle,~R. Multiple
  Charge‐Separation Pathways in Photosystem {II}: Modeling of Transient
  Absorption Kinetics. \emph{Phys. Chem. Chem. Phys.} \textbf{2011}, \emph{12},
  681--688\relax
\mciteBstWouldAddEndPuncttrue
\mciteSetBstMidEndSepPunct{\mcitedefaultmidpunct}
{\mcitedefaultendpunct}{\mcitedefaultseppunct}\relax
\EndOfBibitem
\bibitem[Novoderezhkin \latin{et~al.}(2015)Novoderezhkin, Romero, and van
  Grondelle]{Nov2015}
Novoderezhkin,~V.~I.; Romero,~E.; van Grondelle,~R. How exciton-vibrational
  coherences control charge separation in the photosystem {II} reaction center.
  \emph{Phys. Chem. Chem. Phys.} \textbf{2015}, \emph{17}, 30828--30841\relax
\mciteBstWouldAddEndPuncttrue
\mciteSetBstMidEndSepPunct{\mcitedefaultmidpunct}
{\mcitedefaultendpunct}{\mcitedefaultseppunct}\relax
\EndOfBibitem
\bibitem[Gelzinis \latin{et~al.}(2013)Gelzinis, Valkunas, Fuller, Ogilvie,
  Mukamel, and Abramavicius]{Mukamel2013}
Gelzinis,~A.; Valkunas,~L.; Fuller,~F.~D.; Ogilvie,~J.~P.; Mukamel,~S.;
  Abramavicius,~D. Tight-binding model of the photosystem {II} reaction center:
  application to two-dimensional electronic spectroscopy. \emph{New J. Phys.}
  \textbf{2013}, \emph{15}, 075013\relax
\mciteBstWouldAddEndPuncttrue
\mciteSetBstMidEndSepPunct{\mcitedefaultmidpunct}
{\mcitedefaultendpunct}{\mcitedefaultseppunct}\relax
\EndOfBibitem
\bibitem[Adolphs and Renger(2006)Adolphs, and Renger]{Renger06}
Adolphs,~J.; Renger,~T. How Proteins Trigger Excitation Energy Transfer in the
  {FMO} Complex of Green Sulfur Bacteria. \emph{Biophys. J.} \textbf{2006},
  \emph{91}, 2778--2797\relax
\mciteBstWouldAddEndPuncttrue
\mciteSetBstMidEndSepPunct{\mcitedefaultmidpunct}
{\mcitedefaultendpunct}{\mcitedefaultseppunct}\relax
\EndOfBibitem
\bibitem[Garg \latin{et~al.}(1985)Garg, Onuchic, and Ambegaokar]{Garg1985}
Garg,~A.; Onuchic,~J.~N.; Ambegaokar,~V. Effect of friction on electron
  transfer in biomolecules. \emph{J. Chem. Phys.} \textbf{1985}, \emph{83},
  4491\relax
\mciteBstWouldAddEndPuncttrue
\mciteSetBstMidEndSepPunct{\mcitedefaultmidpunct}
{\mcitedefaultendpunct}{\mcitedefaultseppunct}\relax
\EndOfBibitem
\bibitem[Wolynes(1987)]{Wolynes1987}
Wolynes,~P.~G. Dissipation, tunneling, and adiabaticity criteria for curve
  crossing problems in the condensed phase. \emph{J. Chem. Phys.}
  \textbf{1987}, \emph{86}, 1957\relax
\mciteBstWouldAddEndPuncttrue
\mciteSetBstMidEndSepPunct{\mcitedefaultmidpunct}
{\mcitedefaultendpunct}{\mcitedefaultseppunct}\relax
\EndOfBibitem
\bibitem[Yan \latin{et~al.}(1988)Yan, Sparpaglione, and Mukamel]{Mukamel1988}
Yan,~Y.~J.; Sparpaglione,~M.; Mukamel,~S. Solvation dynamics in
  electron-transfer, isomerization, and nonlinear optical processes: a unified
  Liouville-space theory. \emph{J. Phys. Chem.} \textbf{1988}, \emph{92},
  4842--4853\relax
\mciteBstWouldAddEndPuncttrue
\mciteSetBstMidEndSepPunct{\mcitedefaultmidpunct}
{\mcitedefaultendpunct}{\mcitedefaultseppunct}\relax
\EndOfBibitem
\bibitem[Tanaka and Tanimura(2009)Tanaka, and Tanimura]{TanakaJPSJ09}
Tanaka,~M.; Tanimura,~Y. Quantum Dissipative Dynamics of Electron Transfer
  Reaction System: Nonperturbative Hierarchy Equations Approach. \emph{J. Phys.
  Soc. Jpn.} \textbf{2009}, \emph{78}, 073802\relax
\mciteBstWouldAddEndPuncttrue
\mciteSetBstMidEndSepPunct{\mcitedefaultmidpunct}
{\mcitedefaultendpunct}{\mcitedefaultseppunct}\relax
\EndOfBibitem
\bibitem[Tanaka and Tanimura(2010)Tanaka, and Tanimura]{TanakaJCP10}
Tanaka,~M.; Tanimura,~Y. Multistate electron transfer dynamics in the condensed
  phase: Exact calculations from the reduced hierarchy equations of motion
  approach. \emph{J. Chem. Phys.} \textbf{2010}, \emph{132}, 214502\relax
\mciteBstWouldAddEndPuncttrue
\mciteSetBstMidEndSepPunct{\mcitedefaultmidpunct}
{\mcitedefaultendpunct}{\mcitedefaultseppunct}\relax
\EndOfBibitem
\bibitem[Sim and Makri(1997)Sim, and Makri]{SIM_MAKRI97}
Sim,~E.; Makri,~N. Path Integral Simulation of Charge Transfer Dynamics in
  Photosynthetic Reaction Centers. \emph{J. Phys. Chem. B} \textbf{1997},
  \emph{101}, 5446--5458\relax
\mciteBstWouldAddEndPuncttrue
\mciteSetBstMidEndSepPunct{\mcitedefaultmidpunct}
{\mcitedefaultendpunct}{\mcitedefaultseppunct}\relax
\EndOfBibitem
\bibitem[Sim(2004)]{SIM2004}
Sim,~E. Determination of the Electron Transfer Mechanism through Decomposition
  of the Density Matrix. \emph{J. Phys. Chem. B} \textbf{2004}, \emph{108},
  19093--19095\relax
\mciteBstWouldAddEndPuncttrue
\mciteSetBstMidEndSepPunct{\mcitedefaultmidpunct}
{\mcitedefaultendpunct}{\mcitedefaultseppunct}\relax
\EndOfBibitem
\bibitem[Dijkstra and Tanimura(2010)Dijkstra, and Tanimura]{DijkstraNJP10DNA}
Dijkstra,~A.~G.; Tanimura,~Y. Correlated fluctuations in the exciton dynamics
  and spectroscopy of {DNA}. \emph{New J. Phys.} \textbf{2010}, \emph{12},
  055005\relax
\mciteBstWouldAddEndPuncttrue
\mciteSetBstMidEndSepPunct{\mcitedefaultmidpunct}
{\mcitedefaultendpunct}{\mcitedefaultseppunct}\relax
\EndOfBibitem
\bibitem[Gélinas \latin{et~al.}(2014)Gélinas, Rao, Kumar, Smith, Chin, and
  Clark]{Gelinas2014}
Gélinas,~S.; Rao,~A.; Kumar,~A.; Smith,~S.~L.; Chin,~A.~W.; Clark,~J.
  Ultrafast Long-Range Charge Separation in Organic Semiconductor Photovoltaic
  Diodes. \textbf{2014}, \emph{343}, 6\relax
\mciteBstWouldAddEndPuncttrue
\mciteSetBstMidEndSepPunct{\mcitedefaultmidpunct}
{\mcitedefaultendpunct}{\mcitedefaultseppunct}\relax
\EndOfBibitem
\bibitem[Zirzlmeier \latin{et~al.}(2015)Zirzlmeier, Lehnherr, Coto, Chernick,
  Casillas, Basel, Thoss, Tykwinski, and Guldi]{Thoss2015}
Zirzlmeier,~J.; Lehnherr,~D.; Coto,~P.~B.; Chernick,~E.~T.; Casillas,~R.;
  Basel,~B.~S.; Thoss,~M.; Tykwinski,~R.~R.; Guldi,~D.~M. Singlet fission in
  pentacene dimers. \emph{Proc. Natl. Acad. Sci. U.S.A.} \textbf{2015},
  \emph{112}, 5325--5330\relax
\mciteBstWouldAddEndPuncttrue
\mciteSetBstMidEndSepPunct{\mcitedefaultmidpunct}
{\mcitedefaultendpunct}{\mcitedefaultseppunct}\relax
\EndOfBibitem
\bibitem[Tamura and Burghardt(2013)Tamura, and Burghardt]{Tamura2013}
Tamura,~H.; Burghardt,~I. Ultrafast Charge Separation in Organic Photovoltaics
  Enhanced by Charge Delocalization and Vibronically Hot Exciton Dissociation.
  \emph{J. Am. Chem. Soc.} \textbf{2013}, \emph{135}, 16364--16367\relax
\mciteBstWouldAddEndPuncttrue
\mciteSetBstMidEndSepPunct{\mcitedefaultmidpunct}
{\mcitedefaultendpunct}{\mcitedefaultseppunct}\relax
\EndOfBibitem
\bibitem[Huix-Rotllant \latin{et~al.}(2015)Huix-Rotllant, Tamura, and
  Burghardt]{TamuraJPC2015}
Huix-Rotllant,~M.; Tamura,~H.; Burghardt,~I. Concurrent Effects of
  Delocalization and Internal Conversion Tune Charge Separation at Regioregular
  Polythiophene–Fullerene Heterojunctions. \emph{J. Phys. Chem. Lett.}
  \textbf{2015}, \emph{6}, 1702--1708\relax
\mciteBstWouldAddEndPuncttrue
\mciteSetBstMidEndSepPunct{\mcitedefaultmidpunct}
{\mcitedefaultendpunct}{\mcitedefaultseppunct}\relax
\EndOfBibitem
\bibitem[Oviedo-Casado \latin{et~al.}(2017)Oviedo-Casado, Urbina, and
  Prior]{Prior2017}
Oviedo-Casado,~S.; Urbina,~A.; Prior,~J. Magnetic field enhancement of organic
  photovoltaic cells performance. \emph{Sci. Rep.} \textbf{2017}, \emph{7},
  4297\relax
\mciteBstWouldAddEndPuncttrue
\mciteSetBstMidEndSepPunct{\mcitedefaultmidpunct}
{\mcitedefaultendpunct}{\mcitedefaultseppunct}\relax
\EndOfBibitem
\bibitem[Cainelli and Tanimura(2021)Cainelli, and Tanimura]{Mauro2020}
Cainelli,~M.; Tanimura,~Y. Exciton Transfer in Organic Photovoltaic Cells: A
  Role of Local and Nonlocal Electron-Phonon Interactions in a Donor Domain.
  \emph{J. Chem. Phys.} \textbf{2021}, \emph{154}, 034107\relax
\mciteBstWouldAddEndPuncttrue
\mciteSetBstMidEndSepPunct{\mcitedefaultmidpunct}
{\mcitedefaultendpunct}{\mcitedefaultseppunct}\relax
\EndOfBibitem
\bibitem[Tanimura(2006)]{TanimuraJPSJ06}
Tanimura,~Y. Stochastic Liouville, Langevin, Fokker–Planck, and Master
  Equation Approaches to Quantum Dissipative Systems. \emph{J. Phys. Soc. Jpn.}
  \textbf{2006}, \emph{75}, 082001\relax
\mciteBstWouldAddEndPuncttrue
\mciteSetBstMidEndSepPunct{\mcitedefaultmidpunct}
{\mcitedefaultendpunct}{\mcitedefaultseppunct}\relax
\EndOfBibitem
\bibitem[Tanimura(2020)]{YTpers}
Tanimura,~Y. Numerically “exact” approach to open quantum dynamics: The
  hierarchical equations of motion ({HEOM}). \emph{J. Chem. Phys.}
  \textbf{2020}, \emph{153}, 020901\relax
\mciteBstWouldAddEndPuncttrue
\mciteSetBstMidEndSepPunct{\mcitedefaultmidpunct}
{\mcitedefaultendpunct}{\mcitedefaultseppunct}\relax
\EndOfBibitem
\bibitem[Dral \latin{et~al.}(2018)Dral, Barbatti, and Thiel]{Pavlo2018ML}
Dral,~P.~O.; Barbatti,~M.; Thiel,~W. Nonadiabatic Excited-State Dynamics with
  Machine Learning. \emph{J. Phys. Chem. Lett.} \textbf{2018}, \emph{9},
  5660--5663\relax
\mciteBstWouldAddEndPuncttrue
\mciteSetBstMidEndSepPunct{\mcitedefaultmidpunct}
{\mcitedefaultendpunct}{\mcitedefaultseppunct}\relax
\EndOfBibitem
\bibitem[Hartmann and Carleo(2019)Hartmann, and Carleo]{Hartmann2019NNQD}
Hartmann,~M.~J.; Carleo,~G. Neural-Network Approach to Dissipative Quantum
  Many-Body Dynamics. \emph{Phys. Rev. Lett.} \textbf{2019}, \emph{122},
  250502\relax
\mciteBstWouldAddEndPuncttrue
\mciteSetBstMidEndSepPunct{\mcitedefaultmidpunct}
{\mcitedefaultendpunct}{\mcitedefaultseppunct}\relax
\EndOfBibitem
\bibitem[Flurin \latin{et~al.}(2020)Flurin, Martin, Hacohen-Gourgy, and
  Siddiqi]{Flurin2020RNN}
Flurin,~E.; Martin,~L.; Hacohen-Gourgy,~S.; Siddiqi,~I. Using a Recurrent
  Neural Network to Reconstruct Quantum Dynamics of a Superconducting Qubit
  from Physical Observations. \emph{Phys. Rev. X} \textbf{2020}, \emph{10},
  011006\relax
\mciteBstWouldAddEndPuncttrue
\mciteSetBstMidEndSepPunct{\mcitedefaultmidpunct}
{\mcitedefaultendpunct}{\mcitedefaultseppunct}\relax
\EndOfBibitem
\bibitem[Zheng \latin{et~al.}(2019)Zheng, Gao, and
  Eisfeld]{zheng_excitonic_2019}
Zheng,~F.; Gao,~X.; Eisfeld,~A. Excitonic Wave Function Reconstruction from
  Near-Field Spectra Using Machine Learning Techniques. \textbf{2019},
  \emph{123}, 163202\relax
\mciteBstWouldAddEndPuncttrue
\mciteSetBstMidEndSepPunct{\mcitedefaultmidpunct}
{\mcitedefaultendpunct}{\mcitedefaultseppunct}\relax
\EndOfBibitem
\bibitem[Rodríguez and Kramer(2019)Rodríguez, and
  Kramer]{rodriguez_machine_2019}
Rodríguez,~M.; Kramer,~T. Machine learning of two-dimensional spectroscopic
  data. \textbf{2019}, \emph{520}, 52--60\relax
\mciteBstWouldAddEndPuncttrue
\mciteSetBstMidEndSepPunct{\mcitedefaultmidpunct}
{\mcitedefaultendpunct}{\mcitedefaultseppunct}\relax
\EndOfBibitem
\bibitem[Namuduri \latin{et~al.}(2020)Namuduri, Titze, Bhansali, and
  Li]{namuduri_machine_2020}
Namuduri,~S.; Titze,~M.; Bhansali,~S.; Li,~H. Machine Learning Enabled
  Lineshape Analysis in Optical Two-Dimensional Coherent Spectroscopy.
  \textbf{2020}, \emph{37}, 1587\relax
\mciteBstWouldAddEndPuncttrue
\mciteSetBstMidEndSepPunct{\mcitedefaultmidpunct}
{\mcitedefaultendpunct}{\mcitedefaultseppunct}\relax
\EndOfBibitem
\bibitem[Smith \latin{et~al.}(2017)Smith, Isayev, and Roitberg]{Smith2017ANI1}
Smith,~J.~S.; Isayev,~O.; Roitberg,~A.~E. {ANI}-1: an extensible neural network
  potential with {DFT} accuracy at force field computational cost. \emph{Chem.
  Sci.} \textbf{2017}, \emph{8}, 3192--3203\relax
\mciteBstWouldAddEndPuncttrue
\mciteSetBstMidEndSepPunct{\mcitedefaultmidpunct}
{\mcitedefaultendpunct}{\mcitedefaultseppunct}\relax
\EndOfBibitem
\bibitem[Sauceda \latin{et~al.}(2019)Sauceda, Chmiela, Poltavsky, Müller, and
  Tkatchenko]{MLMD01}
Sauceda,~H.~E.; Chmiela,~S.; Poltavsky,~I.; Müller,~K.-R.; Tkatchenko,~A.
  Molecular force fields with gradient-domain machine learning: Construction
  and application to dynamics of small molecules with coupled cluster forces.
  \emph{J. Chem. Phys.} \textbf{2019}, \emph{150}, 114102\relax
\mciteBstWouldAddEndPuncttrue
\mciteSetBstMidEndSepPunct{\mcitedefaultmidpunct}
{\mcitedefaultendpunct}{\mcitedefaultseppunct}\relax
\EndOfBibitem
\bibitem[Sifain \latin{et~al.}(2018)Sifain, Lubbers, Nebgen, Smith, Lokhov,
  Isayev, Roitberg, Barros, and Tretiak]{MLMD02}
Sifain,~A.~E.; Lubbers,~N.; Nebgen,~B.~T.; Smith,~J.~S.; Lokhov,~A.~Y.;
  Isayev,~O.; Roitberg,~A.~E.; Barros,~K.; Tretiak,~S. Discovering a
  Transferable Charge Assignment Model Using Machine Learning. \emph{J. Phys.
  Chem. Lett.} \textbf{2018}, \emph{9}, 4495--4501\relax
\mciteBstWouldAddEndPuncttrue
\mciteSetBstMidEndSepPunct{\mcitedefaultmidpunct}
{\mcitedefaultendpunct}{\mcitedefaultseppunct}\relax
\EndOfBibitem
\bibitem[Chmiela \latin{et~al.}(2017)Chmiela, Tkatchenko, Sauceda, Poltavsky,
  Schütt, and Müller]{MLMD03}
Chmiela,~S.; Tkatchenko,~A.; Sauceda,~H.~E.; Poltavsky,~I.; Schütt,~K.~T.;
  Müller,~K.-R. Machine learning of accurate energy-conserving molecular force
  fields. \emph{Sci. Adv.} \textbf{2017}, \emph{3}, e1603015\relax
\mciteBstWouldAddEndPuncttrue
\mciteSetBstMidEndSepPunct{\mcitedefaultmidpunct}
{\mcitedefaultendpunct}{\mcitedefaultseppunct}\relax
\EndOfBibitem
\bibitem[Chmiela \latin{et~al.}(2018)Chmiela, Sauceda, Müller, and
  Tkatchenko]{MLMD04}
Chmiela,~S.; Sauceda,~H.~E.; Müller,~K.-R.; Tkatchenko,~A. Towards exact
  molecular dynamics simulations with machine-learned force fields.
  \emph{Nature Comm.} \textbf{2018}, \emph{9}, 1--10\relax
\mciteBstWouldAddEndPuncttrue
\mciteSetBstMidEndSepPunct{\mcitedefaultmidpunct}
{\mcitedefaultendpunct}{\mcitedefaultseppunct}\relax
\EndOfBibitem
\bibitem[Kampen(1981)]{Kampen81}
Kampen,~N.~V. \emph{Stochastic Processes in Physics and Chemistry}; Elsevier,
  1981\relax
\mciteBstWouldAddEndPuncttrue
\mciteSetBstMidEndSepPunct{\mcitedefaultmidpunct}
{\mcitedefaultendpunct}{\mcitedefaultseppunct}\relax
\EndOfBibitem
\bibitem[Chen \latin{et~al.}(2015)Chen, Zhao, and Tanimura]{LipenHolns2015}
Chen,~L.; Zhao,~Y.; Tanimura,~Y. Dynamics of a One-Dimensional Holstein Polaron
  with the Hierarchical Equations of Motion Approach. \emph{J. Phys. Chem.
  Lett.} \textbf{2015}, \emph{6}, 3110--3115\relax
\mciteBstWouldAddEndPuncttrue
\mciteSetBstMidEndSepPunct{\mcitedefaultmidpunct}
{\mcitedefaultendpunct}{\mcitedefaultseppunct}\relax
\EndOfBibitem
\bibitem[Dunn \latin{et~al.}(2019)Dunn, Tempelaar, and
  Reichman]{ReichmanHolns2019}
Dunn,~I.~S.; Tempelaar,~R.; Reichman,~D.~R. Removing instabilities in the
  hierarchical equations of motion: Exact and approximate projection
  approaches. \emph{J. Chem. Phys.} \textbf{2019}, \emph{150}, 184109\relax
\mciteBstWouldAddEndPuncttrue
\mciteSetBstMidEndSepPunct{\mcitedefaultmidpunct}
{\mcitedefaultendpunct}{\mcitedefaultseppunct}\relax
\EndOfBibitem
\bibitem[Ueno and Tanimura(2020)Ueno, and Tanimura]{Ueno2020}
Ueno,~S.; Tanimura,~Y. Modeling intermolecular and intramolecular modes of
  liquid water using multiple heat baths: Machine learning approach. \emph{J.
  Chem. Theory Comput.} \textbf{2020}, \emph{16}, 2099--2108\relax
\mciteBstWouldAddEndPuncttrue
\mciteSetBstMidEndSepPunct{\mcitedefaultmidpunct}
{\mcitedefaultendpunct}{\mcitedefaultseppunct}\relax
\EndOfBibitem
\bibitem[Halpin \latin{et~al.}(2014)Halpin, Johnson, Tempelaar, Murphy,
  Knoester, Jansen, and Miller]{Dwayne2014}
Halpin,~A.; Johnson,~P. J.~M.; Tempelaar,~R.; Murphy,~R.~S.; Knoester,~J.;
  Jansen,~T. L.~C.; Miller,~R. J.~D. Two-dimensional spectroscopy of a
  molecular dimer unveils the effects of vibronic coupling on exciton
  coherences. \emph{Nature Chem.} \textbf{2014}, \emph{6}, 196--201\relax
\mciteBstWouldAddEndPuncttrue
\mciteSetBstMidEndSepPunct{\mcitedefaultmidpunct}
{\mcitedefaultendpunct}{\mcitedefaultseppunct}\relax
\EndOfBibitem
\bibitem[Tempelaar \latin{et~al.}(2016)Tempelaar, Halpin, Johnson, Cai, Murphy,
  Knoester, Miller, and Jansen]{tempelaar_laser-limited_2016}
Tempelaar,~R.; Halpin,~A.; Johnson,~P. J.~M.; Cai,~J.; Murphy,~R.~S.;
  Knoester,~J.; Miller,~R. J.~D.; Jansen,~T. L.~C. Laser-Limited Signatures of
  Quantum Coherence. \emph{The Journal of Physical Chemistry A} \textbf{2016},
  \emph{120}, 3042--3048, PMID: 26558888\relax
\mciteBstWouldAddEndPuncttrue
\mciteSetBstMidEndSepPunct{\mcitedefaultmidpunct}
{\mcitedefaultendpunct}{\mcitedefaultseppunct}\relax
\EndOfBibitem
\bibitem[Duan \latin{et~al.}(2015)Duan, Nalbach, Prokhorenko, Mukamel, and
  Thorwart]{duan_origin_2015}
Duan,~H.-G.; Nalbach,~P.; Prokhorenko,~V.~I.; Mukamel,~S.; Thorwart,~M. On the
  origin of oscillations in two-dimensional spectra of excitonically-coupled
  molecular systems. \emph{New Journal of Physics} \textbf{2015}, \emph{17},
  072002\relax
\mciteBstWouldAddEndPuncttrue
\mciteSetBstMidEndSepPunct{\mcitedefaultmidpunct}
{\mcitedefaultendpunct}{\mcitedefaultseppunct}\relax
\EndOfBibitem
\bibitem[Berendsen \latin{et~al.}(1995)Berendsen, van~der Spoel, and van
  Drunen]{berendsen1995gromacs}
Berendsen,~H.; van~der Spoel,~D.; van Drunen,~R. {GROMACS}: A message-passing
  parallel molecular dynamics implementation. \emph{Comput. Phys. Comm.}
  \textbf{1995}, \emph{91}, 43--56\relax
\mciteBstWouldAddEndPuncttrue
\mciteSetBstMidEndSepPunct{\mcitedefaultmidpunct}
{\mcitedefaultendpunct}{\mcitedefaultseppunct}\relax
\EndOfBibitem
\bibitem[Abraham \latin{et~al.}(2015)Abraham, Murtola, Schulz, Páll, Smith,
  Hess, and Lindahl]{abraham2015gromacs}
Abraham,~M.~J.; Murtola,~T.; Schulz,~R.; Páll,~S.; Smith,~J.~C.; Hess,~B.;
  Lindahl,~E. {GROMACS}: High performance molecular simulations through
  multi-level parallelism from laptops to supercomputers. \emph{{SoftwareX}}
  \textbf{2015}, \emph{1-2}, 19--25\relax
\mciteBstWouldAddEndPuncttrue
\mciteSetBstMidEndSepPunct{\mcitedefaultmidpunct}
{\mcitedefaultendpunct}{\mcitedefaultseppunct}\relax
\EndOfBibitem
\bibitem[Bekker \latin{et~al.}(1993)Bekker, Berendsen, Dijkstra, Achterop,
  Van~Drunen, Van~der Spoel, Sijbers, Keegstra, Reitsma, and
  Renardus]{bekker1993gromacs}
Bekker,~H.; Berendsen,~H.; Dijkstra,~E.; Achterop,~S.; Van~Drunen,~R.; Van~der
  Spoel,~D.; Sijbers,~A.; Keegstra,~H.; Reitsma,~B.; Renardus,~M. Gromacs: A
  parallel computer for molecular dynamics simulations. Physics computing.
  1993; pp 252--256\relax
\mciteBstWouldAddEndPuncttrue
\mciteSetBstMidEndSepPunct{\mcitedefaultmidpunct}
{\mcitedefaultendpunct}{\mcitedefaultseppunct}\relax
\EndOfBibitem
\bibitem[Ridley and Zerner(1973)Ridley, and Zerner]{ZINDO1973}
Ridley,~J.; Zerner,~M. An intermediate neglect of differential overlap
  technique for spectroscopy: Pyrrole and the azines. \emph{Theoret. Chim.
  Acta} \textbf{1973}, \emph{32}, 111--134\relax
\mciteBstWouldAddEndPuncttrue
\mciteSetBstMidEndSepPunct{\mcitedefaultmidpunct}
{\mcitedefaultendpunct}{\mcitedefaultseppunct}\relax
\EndOfBibitem
\bibitem[Thompson and Zerner(1991)Thompson, and Zerner]{ZINDO1991}
Thompson,~M.~A.; Zerner,~M.~C. A theoretical examination of the electronic
  structure and spectroscopy of the photosynthetic reaction center from
  Rhodopseudomonas viridis. \emph{J. Am. Chem. Soc.} \textbf{1991}, \emph{113},
  8210--8215\relax
\mciteBstWouldAddEndPuncttrue
\mciteSetBstMidEndSepPunct{\mcitedefaultmidpunct}
{\mcitedefaultendpunct}{\mcitedefaultseppunct}\relax
\EndOfBibitem
\bibitem[Martin(2003)]{NTO2003}
Martin,~R.~L. Natural transition orbitals. \emph{J. Chem. Phys.} \textbf{2003},
  \emph{118}, 4775\relax
\mciteBstWouldAddEndPuncttrue
\mciteSetBstMidEndSepPunct{\mcitedefaultmidpunct}
{\mcitedefaultendpunct}{\mcitedefaultseppunct}\relax
\EndOfBibitem
\bibitem[Neese(2012)]{neese2012orca}
Neese,~F. The {ORCA} program system. \emph{{WIREs} Comput. Mol. Sci.}
  \textbf{2012}, \emph{2}, 73--78\relax
\mciteBstWouldAddEndPuncttrue
\mciteSetBstMidEndSepPunct{\mcitedefaultmidpunct}
{\mcitedefaultendpunct}{\mcitedefaultseppunct}\relax
\EndOfBibitem
\bibitem[Abadi \latin{et~al.}(2016)Abadi, Agarwal, Barham, Brevdo, Chen, Citro,
  Corrado, Davis, Dean, Devin, Ghemawat, Goodfellow, Harp, Irving, Isard, Jia,
  Jozefowicz, Kaiser, Kudlur, Levenberg, Mane, Monga, Moore, Murray, Olah,
  Schuster, Shlens, Steiner, Sutskever, Talwar, Tucker, Vanhoucke, Vasudevan,
  Viegas, Vinyals, Warden, Wattenberg, Wicke, Yu, and Zheng]{tensorflow}
Abadi,~M.; Agarwal,~A.; Barham,~P.; Brevdo,~E.; Chen,~Z.; Citro,~C.;
  Corrado,~G.~S.; Davis,~A.; Dean,~J.; Devin,~M.; Ghemawat,~S.; Goodfellow,~I.;
  Harp,~A.; Irving,~G.; Isard,~M.; Jia,~Y.; Jozefowicz,~R.; Kaiser,~L.;
  Kudlur,~M.; Levenberg,~J.; Mane,~D.; Monga,~R.; Moore,~S.; Murray,~D.;
  Olah,~C.; Schuster,~M.; Shlens,~J.; Steiner,~B.; Sutskever,~I.; Talwar,~K.;
  Tucker,~P.; Vanhoucke,~V.; Vasudevan,~V.; Viegas,~F.; Vinyals,~O.;
  Warden,~P.; Wattenberg,~M.; Wicke,~M.; Yu,~Y.; Zheng,~X. {TensorFlow}:
  Large-Scale Machine Learning on Heterogeneous Distributed Systems.
  \emph{{arXiv}:1603.04467 [cs]} \textbf{2016}, \relax
\mciteBstWouldAddEndPunctfalse
\mciteSetBstMidEndSepPunct{\mcitedefaultmidpunct}
{}{\mcitedefaultseppunct}\relax
\EndOfBibitem
\bibitem[Tanimura(2012)]{TanimruaJCP12}
Tanimura,~Y. Reduced hierarchy equations of motion approach with Drude plus
  Brownian spectral distribution: Probing electron transfer processes by means
  of two-dimensional correlation spectroscopy. \emph{J. Chem. Phys.}
  \textbf{2012}, \emph{137}, 22A550\relax
\mciteBstWouldAddEndPuncttrue
\mciteSetBstMidEndSepPunct{\mcitedefaultmidpunct}
{\mcitedefaultendpunct}{\mcitedefaultseppunct}\relax
\EndOfBibitem
\bibitem[Tanimura and Mukamel(1995)Tanimura, and Mukamel]{TanimuraMukamelJCP95}
Tanimura,~Y.; Mukamel,~S. Femtosecond pump–probe spectroscopy of
  intermolecular vibrations in molecular dimers. \emph{J. Chem. Phys.}
  \textbf{1995}, \emph{103}, 1981--1984\relax
\mciteBstWouldAddEndPuncttrue
\mciteSetBstMidEndSepPunct{\mcitedefaultmidpunct}
{\mcitedefaultendpunct}{\mcitedefaultseppunct}\relax
\EndOfBibitem
\bibitem[Tanimura(2014)]{Y.Tanimura.JCP.2014}
Tanimura,~Y. Reduced hierarchical equations of motion in real and imaginary
  time: Correlated initial states and thermodynamic quantities. \emph{The
  Journal of Chemical Physics} \textbf{2014}, \emph{141}, 044114\relax
\mciteBstWouldAddEndPuncttrue
\mciteSetBstMidEndSepPunct{\mcitedefaultmidpunct}
{\mcitedefaultendpunct}{\mcitedefaultseppunct}\relax
\EndOfBibitem
\bibitem[Tanimura(2015)]{Y.Tanimura.JCP.2015}
Tanimura,~Y. Real-time and imaginary-time quantum hierarchal Fokker-Planck
  equations. \emph{The Journal of Chemical Physics} \textbf{2015}, \emph{142},
  144110\relax
\mciteBstWouldAddEndPuncttrue
\mciteSetBstMidEndSepPunct{\mcitedefaultmidpunct}
{\mcitedefaultendpunct}{\mcitedefaultseppunct}\relax
\EndOfBibitem
\bibitem[Ikeda and Tanimura(2017)Ikeda, and Tanimura]{T.Ikeda.JCP.2017}
Ikeda,~T.; Tanimura,~Y. Probing photoisomerization processes by means of
  multi-dimensional electronic spectroscopy: The multi-state quantum
  hierarchical Fokker-Planck equation approach. \emph{The Journal of Chemical
  Physics} \textbf{2017}, \emph{147}, 014102\relax
\mciteBstWouldAddEndPuncttrue
\mciteSetBstMidEndSepPunct{\mcitedefaultmidpunct}
{\mcitedefaultendpunct}{\mcitedefaultseppunct}\relax
\EndOfBibitem
\bibitem[Ikeda \latin{et~al.}(2019)Ikeda, Dijkstra, and Tanimura]{Ikeda2019}
Ikeda,~T.; Dijkstra,~A.; Tanimura,~Y. Modeling and analyzing a photo-driven
  molecular motor system: Ratchet dynamics and non-linear optical spectra.
  \emph{J. Chem. Phys.} \textbf{2019}, \emph{150}, 114103\relax
\mciteBstWouldAddEndPuncttrue
\mciteSetBstMidEndSepPunct{\mcitedefaultmidpunct}
{\mcitedefaultendpunct}{\mcitedefaultseppunct}\relax
\EndOfBibitem
\bibitem[Ikeda and Tanimura(2018)Ikeda, and Tanimura]{Ikeda2018}
Ikeda,~T.; Tanimura,~Y. Phase-space wavepacket dynamics of internal conversion
  via conical intersection: Multi-state quantum Fokker-Planck equation
  approach. \emph{Chem. Phys.} \textbf{2018}, \emph{515}, 203\relax
\mciteBstWouldAddEndPuncttrue
\mciteSetBstMidEndSepPunct{\mcitedefaultmidpunct}
{\mcitedefaultendpunct}{\mcitedefaultseppunct}\relax
\EndOfBibitem
\bibitem[Chen and Shi(2009)Chen, and Shi]{Shi2009PT}
Chen,~L.; Shi,~Q. Quantum rate dynamics for proton transfer reactions in
  condensed phase: The exact hierarchical equations of motion approach.
  \emph{J. Chem. Phys.} \textbf{2009}, \emph{130}, 134505\relax
\mciteBstWouldAddEndPuncttrue
\mciteSetBstMidEndSepPunct{\mcitedefaultmidpunct}
{\mcitedefaultendpunct}{\mcitedefaultseppunct}\relax
\EndOfBibitem
\bibitem[Shi \latin{et~al.}(2011)Shi, Zhu, and Chen]{Shi2011PT}
Shi,~Q.; Zhu,~L.; Chen,~L. Quantum rate dynamics for proton transfer reaction
  in a model system: Effect of the rate promoting vibrational mode. \emph{J.
  Chem. Phys.} \textbf{2011}, \emph{135}, 044505\relax
\mciteBstWouldAddEndPuncttrue
\mciteSetBstMidEndSepPunct{\mcitedefaultmidpunct}
{\mcitedefaultendpunct}{\mcitedefaultseppunct}\relax
\EndOfBibitem
\bibitem[Zhang \latin{et~al.}(2020)Zhang, Borrelli, and Tanimura]{Jianji2020}
Zhang,~J.; Borrelli,~R.; Tanimura,~Y. Proton tunneling in a two-dimensional
  potential energy surface with a non-linear system–bath interaction: Thermal
  suppression of reaction rate. \emph{J. Chem. Phys.} \textbf{2020},
  \emph{152}, 214114\relax
\mciteBstWouldAddEndPuncttrue
\mciteSetBstMidEndSepPunct{\mcitedefaultmidpunct}
{\mcitedefaultendpunct}{\mcitedefaultseppunct}\relax
\EndOfBibitem
\bibitem[Sakamoto and Tanimura(2017)Sakamoto, and Tanimura]{Sakamoto2017}
Sakamoto,~S.; Tanimura,~Y. Exciton-Coupled Electron Transfer Process Controlled
  by Non-Markovian Environments. \emph{J. Phys. Chem. Lett.} \textbf{2017},
  \emph{8}, 5390--5394\relax
\mciteBstWouldAddEndPuncttrue
\mciteSetBstMidEndSepPunct{\mcitedefaultmidpunct}
{\mcitedefaultendpunct}{\mcitedefaultseppunct}\relax
\EndOfBibitem
\bibitem[Zhang \latin{et~al.}(2021)Zhang, Borrelli, and Tanimura]{Jianji2021}
Zhang,~J.; Borrelli,~R.; Tanimura,~Y. Probing photoinduced proton coupled
  electron transfer process by means of two-dimensional electronic-vibrational
  spectroscopy. \emph{J. Chem. Phys.} \textbf{2021}, \emph{15?}, xxx\relax
\mciteBstWouldAddEndPuncttrue
\mciteSetBstMidEndSepPunct{\mcitedefaultmidpunct}
{\mcitedefaultendpunct}{\mcitedefaultseppunct}\relax
\EndOfBibitem
\bibitem[Mukamel(1999)]{mukamel1999book}
Mukamel,~S. \emph{Principles of Nonlinear Optical Spectroscopy}; Oxford
  University Press, 1999\relax
\mciteBstWouldAddEndPuncttrue
\mciteSetBstMidEndSepPunct{\mcitedefaultmidpunct}
{\mcitedefaultendpunct}{\mcitedefaultseppunct}\relax
\EndOfBibitem
\bibitem[Tanimura and Mukamel(1993)Tanimura, and Mukamel]{TanimuraMukamelPRE}
Tanimura,~Y.; Mukamel,~S. Real-time path-integral approach to quantum coherence
  and dephasing in nonadiabatic transitions and nonlinear optical response.
  \emph{Phys. Rev. E} \textbf{1993}, \emph{47}, 118--136\relax
\mciteBstWouldAddEndPuncttrue
\mciteSetBstMidEndSepPunct{\mcitedefaultmidpunct}
{\mcitedefaultendpunct}{\mcitedefaultseppunct}\relax
\EndOfBibitem
\bibitem[Bishop(2006)]{bishop_pattern_2006}
Bishop,~C.~M. \emph{Pattern recognition and machine learning}; Information
  science and statistics; Springer, 2006\relax
\mciteBstWouldAddEndPuncttrue
\mciteSetBstMidEndSepPunct{\mcitedefaultmidpunct}
{\mcitedefaultendpunct}{\mcitedefaultseppunct}\relax
\EndOfBibitem
\end{mcitethebibliography}
\end{document}